\documentclass{article}

\usepackage{arxiv}

\usepackage[utf8]{inputenc} 
\usepackage[T1]{fontenc}    
\usepackage{hyperref}       
\usepackage{url}            
\usepackage{booktabs}       
\usepackage{amsfonts}       
\usepackage{nicefrac}       
\usepackage{microtype}      
\usepackage{lipsum}
\usepackage{natbib}
\usepackage{graphicx} \usepackage{bmpsize}
\usepackage{amsmath}
\usepackage{amsthm}
\usepackage{amssymb}
\usepackage{siunitx}
\usepackage{float}
\usepackage{booktabs}
\usepackage{bm}
\usepackage{todonotes}
\usepackage{hyperref}
\usepackage{caption}
\usepackage{subcaption}
\usepackage{soul}
\usepackage{cancel}
\PassOptionsToPackage{numbers}{natbib}
\usepackage{color}
\usepackage{amsfonts}  
\usepackage{nicefrac} 
\usepackage{tikz}
\usepackage{algorithm}
\usepackage{algpseudocode}

\usepackage{algorithmicx}

\newcommand{\bas}[1]{\begin{align*}#1\end{align*}}
\newcommand{{\phivmat}}{\boldsymbol{{\phi}}}
\newcommand{{\phiv}}{\boldsymbol{{\phi}}}
\newcommand{\ba}[1]{\begin{align}#1\end{align}}

\title{SimCD: Simultaneous Clustering and Differential expression analysis for single-cell transcriptomic data}

\author{
  Seyednami~Niyakan \\
  Department of Electrical \& Computer Engineering\\
  Texas A\&M University\\
  College Station, TX 77843 \\
  \texttt{naminiyakan@tamu.edu} \\
   \And
 Ehsan~Hajiramezanali \\
  Department of Electrical \& Computer Engineering\\
  Texas A\&M University\\
  College Station, TX 77843\\
  \texttt{ehsanr@tamu.edu} \\
   \AND
   Shahin Boluki \\
    Department of Electrical \& Computer Engineering\\
   Texas A\&M University\\
   College Station, TX 77843\\
   \texttt{s.boluki@tamu.edu} \\
   \And
   Siamak Zamani Dadaneh \\
    Department of Electrical \& Computer Engineering\\
    Texas A\&M University\\
    College Station, TX 77843\\
   \texttt{siamak@tamu.edu} \\
   \And
   Xiaoning Qian \\
   Departments of Electrical \& Computer Engineering,\\ Computer Science \& Engineering\\
   Texas A\&M University\\
   College Station, TX 77843\\
   \texttt{xqian@ece.tamu.edu} \\
}

\begin{document}
\maketitle

\begin{abstract}
Single-Cell RNA sequencing (scRNA-seq) measurements have facilitated genome-scale transcriptomic profiling of individual cells, with the hope of deconvolving cellular dynamic changes in corresponding cell sub-populations to better understand molecular mechanisms of different development processes. Several scRNA-seq analysis methods have been proposed to first identify cell sub-populations by clustering and then separately perform differential expression analysis to understand gene expression changes. Their corresponding statistical models and inference algorithms are often designed disjointly. We develop a new method---SimCD---that explicitly models cell heterogeneity and dynamic differential changes in one unified hierarchical gamma-negative binomial (hGNB) model, allowing simultaneous cell clustering and differential expression analysis for scRNA-seq data. Our method naturally defines cell heterogeneity by dynamic expression changes, which is expected to help achieve better performances on the two tasks compared to the existing methods that perform them separately. In addition, SimCD better models dropout (zero inflation) in scRNA-seq data by both cell- and gene-level factors and obviates the need for sophisticated pre-processing steps such as normalization, thanks to the direct modeling of scRNA-seq count data by the rigorous hGNB model with an efficient Gibbs sampling inference algorithm. Extensive comparisons with the state-of-the-art methods on  both simulated and real-world scRNA-seq count data demonstrate the capability of SimCD to discover cell clusters and capture dynamic expression changes. Furthermore, SimCD helps identify several known genes affected by food deprivation in hypothalamic neuron cell subtypes as well as some new potential markers, suggesting the capability of SimCD for bio-marker discovery. SimCD is implemented in R and is available at \url{https://github.com/namini94/SimCD}
\end{abstract}


\section{Introduction}
Recent advances in single-cell RNA sequencing (scRNA-seq) provide great opportunities for enhancing our knowledge of the dynamic cellular processes and characterizing heterogeneity of cell types in many complex tissues \cite{mousehypo2017,paracrine2014,Zeisel2015}. It is challenging to model scRNA-seq count data due to higher levels of both technical and biological noise, highly over-dispersed nature, high-dimensionality, and heterogeneity of the gene expression processes at the single-cell level \cite{jointDRclustering,Mou2020}. A large number of statistical tools have been developed to analyze the gene expression considering the inherent complexity of scRNA-seq data~\cite{lopez2018deep,zinbwave2018,DEsingle,sigEMD}.

Cell-to-cell variations in depth of sequencing and excessive number of zeros in transcriptional profiles from scRNA-seq techniques make downstream analyses difficult. Most of the existing scRNA-seq analysis methods employ common pre-processing steps such as normalization to address the sequencing depth variability across samples~\cite{Lytal2020}. Such pre-processing steps make the performance of analysis depend on suitability of the introduced pre-conditioning for the structure of the scRNA-seq data under the study~\cite{Zyprych2015}. Besides that, different methods have been proposed to deal with the zero inflation property of scRNA-seq data. One way is using imputation methods to replace the zero counts with non-zero values, assuming that technical factors cause the zero inflation~\cite{DrImpute2018}. Another broadly used technique is to explicitly model the count distribution with a negative binomial~(NB) distributed random variable with a zero-inflated~(ZI) component that generates zeros (known as the ZINB distribution)~\cite{zinbwave2018}. However, recent works have shown that using imputation routines or ZINB models is unnecessary and may destroy the underlying biological signal in scRNA-seq data~\cite{choiK2020zerosorigin}. On the other hand, using negative binomial based models that account for known biological confounding factors, such as cell types, treatment conditions, and sex, can better model zero inflation~\cite{choiK2020zerosorigin}.

A common practice in scRNA-seq analysis is to identify cell sub-populations by clustering algorithms, then followed by separate differential expression analysis between detected cell clusters to discover cluster-specific marker genes. Most of the existing cell clustering and differential expression analysis methods are only designed to carry out one of these two tasks \cite{sigEMD,DEsingle,deseq2,zinbwave2018}. However, a unified distributional model that can simultaneously perform both clustering and differential expression analysis assures more consistent results. Recently, a new hierarchical Bayesian model, scVI~\cite{lopez2018deep}, has been developed to perform both clustering and differential expression analysis. 
Several trajectory-based methods have also been proposed for differential expression analysis both within and across conditions based on (pseudo)-dynamic changes considering the underlying dynamical biological processes \cite{van2020trajectory,campbell2018uncovering}. 

Here, we introduce, SimCD, a unified Bayesian method based on a hierarchical gamma-negative binomial (hGNB) model, to simultaneously perform clustering and differential expression analysis. With an efficient inference algorithm, SimCD infers the gene and cell specific parameters that are designed to inherently model the sample heterogeneity and dynamic gene expression changes so that resulting cell sub-populations by clustering can capture dynamic expression changes. SimCD is capable of including both gene- and cell-level biological explanatory variables to better model zero inflation in scRNA-seq data. More critically, SimCD enables dynamic differential expression analysis considering cell heterogeneity for scRNA-seq data across different conditions (for example, phenotypes or treatment conditions). We note that the commonly adopted pre-processing step, surrogate variable analysis (SVA)~\cite{sva2012}, is not needed for scRNA-seq data analysis with SimCD due to its direct modeling of impacts from covariates, which also obviates the need of other pre-processing steps including normalization or count data transformations. 

To demonstrate the capability of SimCD for both cell sub-population identification and differential expression analyses, we have applied SimCD to multiple synthetic datasets of varying characteristics and show that it outperforms the popular state-of-the-art~(SOTA) methods. Furthermore, applying SimCD to real-world scRNA-seq data has showcased its utility of identifying biologically meaningful markers in corresponding molecular mechanisms across different conditions.

\section{Methods}
\noindent {\bf Notations. \quad} Throughout this paper, we use the NB distribution to model scRNA-seq read counts. We parameterize a NB random variable as $Y \sim \text{NB}(r,p)$, where $r$ is the nonnegative dispersion and $p$ is the probability parameter. The probability mass function (PMF) of the random count $Y$ is expressed as $f_Y(y)=Pr(Y=y) = \frac{\Gamma(y+r)}{y!\Gamma(r)}p^y(1-p)^r$, where $\Gamma(\cdot)$ is the gamma function. The NB distribution $Y \sim \text{NB}(r,p)$ can be generated from a compound Poisson distribution:  
\bas{\small
Y = \sum_{t=1}^{L} U_t, \;\; U_t \sim \text{Log}(p), \;\; L \sim \text{Pois}(-r\ln (1-p)), \nonumber
}
where $U\sim\text{Log}(p)$ corresponds to the logarithmic random variable~\citep{johnson2005univariate}, with a PMF $f_U(u) = -\frac{p^u}{u\ln(1-p)}$, $u=1,2,\dots$. As shown in \cite{zhou2015negative}, given $y$ and $r$, the random count $L$ follows a Chinese Restaurant Table (CRT) distribution, $(L\, | \, y,r) \sim \text{CRT}(y,r)$, which can be generated as 
$L = \sum_{t=1}^{y} B_t,~B_t \sim \text{Bernoulli}(\frac{r}{r+t-1})$.

\subsection{Hierarchical gamma-negative binomial (hGNB) model} 

The hierarchical gamma-negative binomial (hGNB) model was recently introduced in \cite{dadaneh2020bayesian} for factor analysis of scRNA-seq count data. More precisely, for the scRNA-seq reads mapped to gene $g$, the read count of a given cell $j$ under different conditions follows the NB distribution: 
$y_{gj} \sim \mbox{NB}(r_j, p_{gj})$.
To handle high variability between different cells, hGNB imposes a gamma prior on the cell-level dispersion parameters: 
	$r_j \sim \mbox{Gamma}(a_0,1/\nu)$,
where $a_0$ and $\nu$ are the shape and rate parameters of the gamma distribution, respectively. We argue that this hierarchical prior structure increases the expressive power of the NB distribution, in particular to better capture potential high over-dispersion observed in scRNA-seq counts. This is unlike most of the existing methods that try to model it using an explicit zero-inflation modeling.

To cluster different cells of scRNA-seq, hGNB uses a latent factor representation model on the logit of the NB probability parameter as 
\ba{\small
	\psi_{gj} = \mbox{logit}(p_{gj}) = \boldsymbol{\phi}_g^T\boldsymbol{\theta}_j,
	\label{eq:p}
}
where the factor loading parameter $\boldsymbol{\phi}_g \in \mathbb{R}^{K \times 1}$ quantifies the association between gene $g$ and latent factor $k$, and the score parameter $\boldsymbol{\theta}_j \in \mathbb{R}^{K \times 1}$ captures the popularity of factor $k$ in cell $j$. The latent factor loading $\boldsymbol{\phi}_g$ and the score parameter $\boldsymbol{\theta}_j$ are assumed to follow an independent Normal distribution:
\ba{\small
\boldsymbol{\phi}_g \sim \mbox{N}(\boldsymbol{\phi}_g;0,I_K),\qquad 
\boldsymbol{\theta}_j \sim \prod_{k=1}^{K} \mbox{N}(\theta_{jk};0,\gamma_k^{-1}).
\label{eq:fac}
}

{To complete the model, hGNB imposes a gamma prior on the rate parameter of gamma distributions, i.e. $\nu$, and also the precision parameters of $\boldsymbol{\theta}_j$. Specifically, throughout the experiments, we set both
the shape ($e_0$) and rate ($f_0$) of these gamma priors to small values ($e_0 = f_0 = 0.01$).}

\subsection{SimCD}

In this paper, we further increase the expressive power of hGNB to derive SimCD for simultaneous clustering and differential expression analysis, by explicitly modeling cell heterogeneity due to different factors. In SimCD, we impose a regression model on the logit of the NB probability parameter:
\ba{\small
	\psi_{gj} = \mbox{logit}(p_{gj}) = \boldsymbol{\phi}_g^T\boldsymbol{\theta}_j + \sum_{n} x_{nj}^{(1)} \beta_{gn}^{(1)} + \sum_{m} x_{mg}^{(2)} \beta_{jm}^{(2)}.
	\label{eq:simcd}
}
While the first term is similar as the one in hGNB and takes clustering into account, $x_{nj}^{(1)}$ and $\beta_{gn}^{(1)}$ in the second term are the design matrix elements and regression coefficients, respectively. More specifically, the second term of the regression model in SimCD helps simultaneous differential  expression analysis. In the simplest case of a comparison across two phenotypes or treatment conditions, the design matrix elements indicate whether a cell $j$ is treated or not, and the regression coefficients adjust the overall expression strength of the gene $g$. Please note that the regression model in simCD provides the flexibility to analyze more complex experiment designs, for instance, differential expression with multiple confounding factors or trajectory-based analysis considering (pseudo)-dynamic changes similar as~\cite{van2020trajectory}. As an example, the covariate coefficients $\beta_{gn}^{(1)}$ can represent variations of interest, such as cell types, or unwanted variations, such as batch effects or quality control measures.
Additionally, $x_{mg}^{(2)}$ are covariates for gene $g$, representing gene length or GC-content for example \citep{risso2011gc}, and $\beta_{jm}^{(2)}$ are their associated regression coefficients. We also include a fixed intercept element in $x_{0g}^{(2)}$ to account for cell-specific expressions, such as the size factors representing differences in sequencing depth.

Both coefficient vectors $\boldsymbol{\beta}_{g}^{(1)}$ and $\boldsymbol{\beta}_{j}^{(2)}$ are assumed to follow the automatic relevance determination (ARD) priors as

\ba{\small
	&\mbox{ARD}(\boldsymbol{\beta}_g^{(1)} \,|\, \alpha_0, \eta_0) = \prod_{n} p(\beta_{gn}^{(1)} \,|\, \alpha_n^{-1}) \,\, p( \alpha_n^{-1} \,|\, \alpha_0, \eta_0),& \nonumber\\
	&\mbox{ARD}(\boldsymbol{\beta}_j^{(2)} \,|\, \alpha_0, \eta_0) = \prod_{m} p(\beta_{jm}^{(2)} \,|\, \eta_m^{-1}) \,\, p( \eta_m^{-1} \,|\, \alpha_0, \eta_0) \nonumber\\
        &
        \beta_{gn}^{(1)} \sim \mbox{Normal}(0,\alpha_n^{-1}), \quad \beta_{jm}^{(2)} \sim \mbox{Normal}(0,\eta_m^{-1}), \nonumber\\
        & 
        \alpha_n \sim \mbox{Gamma}(\alpha_0, \eta_0), \qquad \eta_m \sim \mbox{Gamma}(\alpha_0, \eta_0). \label{eq:reg} 
}
The hyper-parameters  $\alpha_0$ and $\eta_0$ are set to small values ($\alpha_0 = \eta_0 = 0.01$ in our experiments) to obtain a  non-informative prior with wide support~\citep{klami2013bayesian}. Note that the components of the regression coefficients, i.e. $\alpha_n$ and $\alpha_m$, are shared between different genes and cells, respectively; thereby making statistical inference more robust by sharing statistical strengths across genes or samples.

Algorithm~1 summarizes the model inference procedure for SimCD. The detailed Gibbs sampling based updates for the SimCD model can be found in Appendix A.1, where several augmentation techniques are adopted to help achieve efficient model inference. 

\begin{algorithm}[t]
\renewcommand{\thealgorithm}{}
 	\caption{1. SimCD model inference}\label{alg:gibbs}
 	\textbf{Inputs}: scRNA-seq counts, design matrix of covariate effects, $N$\\
 	\textbf{Outputs}: SimCD model parameters following corresponding posteriors 
 	\begin{algorithmic}[1]
 		\State \textit{Initialize} model parameters
 		\State \# Do Gibbs sampling:
 		\For {$iter=1$ to $N$}
 		\State $\,\,$Sample $\ell_{gj}$ using the CRT distribution 
 		\State $\,\,$Update $r_{j}$ using the gamma-Poisson conjugacy 
 		\State $\,\,$Sample auxiliary variables $\omega_{gj}$, using the Polya-Gamma 
 		 distribution 
 		\State $\,\,$Update cell- and gene-level regression coefficients 
		\State $\,\,$Update factor loadings and scores: $\boldsymbol{\phi}_g$ and $\boldsymbol{\theta}_j$ 
		\State $\,\,$Update $\boldsymbol{\alpha}_n$ and $\boldsymbol{\eta}_m$ 
		\State Update $\boldsymbol{\gamma}_k$
		\State Update $\boldsymbol{\nu}$
 		\EndFor 
 	\end{algorithmic}
\end{algorithm}

\subsection{Clustering and Differential expression (DE) analysis}

With the inferred posterior distributions of model parameters in SimCD, we take the inferred posteriors of $\boldsymbol{\theta}_j$ and $\boldsymbol{\beta}_{g}^{(1)}$ for cell clustering and the differential gene expression analysis, respectively. 

To cluster cells, we consider the latent factor representation of the count $y_{gj}$, $\theta_{jk}$ in~\eqref{eq:fac} that captures the popularity of factor $k$ in cell $j$. Specifically, the resampling-based sequential ensemble clustering (RSEC) method \citep{cluster2017} is applied to the inferred score parameters $\theta_{jk}$ to get cell clustering assignments. We follow the workflow explained in \cite{Perraudeau2017} for choices of parameters in the RSEC framework.

In SimCD, since in the prior,
\ba{\small
    \mathbb{E}[y_{gj}] = r_j \,\, \mbox{exp}\left(\boldsymbol{\phi}_g^T\boldsymbol{\theta}_j + \sum_{n} x_{nj}^{(1)} \beta_{gn}^{(1)} + \sum_{m} x_{mg}^{(2)} \beta_{jm}^{(2)}\right); 
}
for the conditional posterior, we have
\ba{\small
    &\mathbb{E}[r_j \,|\, - ] =  \frac{a_0+ \sum_g \ell_{gj}}{\nu + \sum_g \mbox{ln} \left(1 + \mbox{exp} \left(\boldsymbol{\phi}_g^T\boldsymbol{\theta}_j + \sum_{n} x_{nj}^{(1)} \beta_{gn}^{(1)} + \sum_{m} x_{mg}^{(2)} \beta_{jm}^{(2)}\right) \right)}. 
}

Therefore, the NB sample-specific dispersion parameter $r_j$, which depends on all the gene counts of sample $j$ through latent counts $\ell_{gj}$, can help model the sequencing depth of sample 
$j$. One may compare the posterior distributions of the quantity $\mbox{exp}(\sum_{n} x_{nj}^{(1)} \beta_{gn}^{(1)})$ of the same gene across different conditions to assess differential expression of that gene. 

Specifically, to assess whether a certain experimental factor $n$ causes significant expression differences across samples for gene $g$, we collect posterior Markov chain Monte Carlo~(MCMC) samples for the regression coefficient vector $\boldsymbol{\beta}_{g}^{(1)}$ and use these MCMC samples to measure the distance between the posterior distributions of $\mbox{exp}(\beta_{g0}^{(1)})$ and $\mbox{exp}(\beta_{g0}^{(1)} + \beta_{gn}^{(1)})$. More precisely, we use the symmetric Kullback–Leibler~(KL) divergence defined between two discrete distributions. Following \cite{dadaneh2018bnp}, we construct a discrete probability vector for each group of collected MCMC samples, referred to as $\boldsymbol{\pi}$ and $\boldsymbol{\pi}^{\prime}$ for the first and second groups under comparison, respectively. Then, we calculate the symmetric KL-divergence as
\ba{\small
\mbox{KL}\left(\boldsymbol{\pi}, \boldsymbol{\pi}^{\prime} \right) = \sum_{v=1}^V \left( \pi_{v} - \pi^{\prime}_v\right) \mbox{log} \frac{\pi_{v} + \epsilon}{\pi^{\prime}_v + \epsilon},
}
where $\epsilon$ as in \cite{tDMI} is a small constant and we set it to $10^{-10}$ through this paper.


\section{Results \& Discussion}
To evaluate our SimCD method for simultaneous clustering and differential expression~(DE) analysis, we compare its performance with the existing state-of-the-arts~(SOTAs) on both synthetic and real-world scRNA-seq data. In particular, we compare its clustering performance with those of scVI~\citep{lopez2018deep} and ZINB-WaVE~\citep{zinbwave2018}. We also benchmark its DE analysis performance with DESeq2~\citep{deseq2}, DEsingle, sigEMD and scVI, which are four popular DE analysis tools for scRNA-seq data. We note that our SimCD performs clustering and differential expression analysis simultaneously while most of the competing methods, such as DEsingle, sigEMD, DESeq2 and ZINB-WaVE, are SOTAs designed to optimize for clustering or DE specifically. 

We first consider synthetic scRNA-seq data and show that SimCD outperforms the SOTAs in terms of both clustering and DE analysis performance. We then benchmark performances of SimCD with the SOTAs on multiple real-world scRNA-seq datasets of varying sizes and show its capability in revealing biological insights in real-world data. Furthermore, we present a case study on single cells dissociated from adult mouse hypothalamus, revealing biological implications by performing simultaneous clustering and DE analysis using SimCD. We show that SimCD identifies several known genes involved in dietary behavior and also new potential bio-markers affected by food deprivation in hypothalamic neuron cell subtypes. 

Our experiments are performed on a single cluster node with Intel Xeon E5-2680 v4 2.40GHz processor, where it takes around 9 hours for SimCD with 2000 MCMC iterations on a simulated dataset having 10000 genes and 100 samples.

\subsection{Synthetic data}

For comprehensive performance evaluation, we have generated synthetic data with two different generative models: the hGNB model and the zero-inflated NB~(ZINB) distribution from the ZINB-WaVE method in~\cite{zinbwave2018}. For each setting, to make the synthetic data closely resemble real-world scRNA-seq data, we first infer the parameters of the corresponding model based on the mouse hypothalamus scRNA-seq dataset~\citep{mousehypo2017}, and then generate synthetic sequencing counts using the inferred model parameters. Following the instruction from DESeq2 in~\cite{deseq2}, we generate count data for 10000 genes across two conditions, each of which has fifty replicate samples. We randomly select 10\% of genes to be differentially expressed across two conditions. For each generative model we change the corresponding model parameters to simulate the cell clustering structures. 

\subsubsection{Comparison on data simulated by the hGNB generative model}

\begin{figure*}[t!]
    \centering
    \begin{subfigure}{.53\textwidth}
        \includegraphics[width=\textwidth,keepaspectratio, trim=5 5 10 10, clip ]{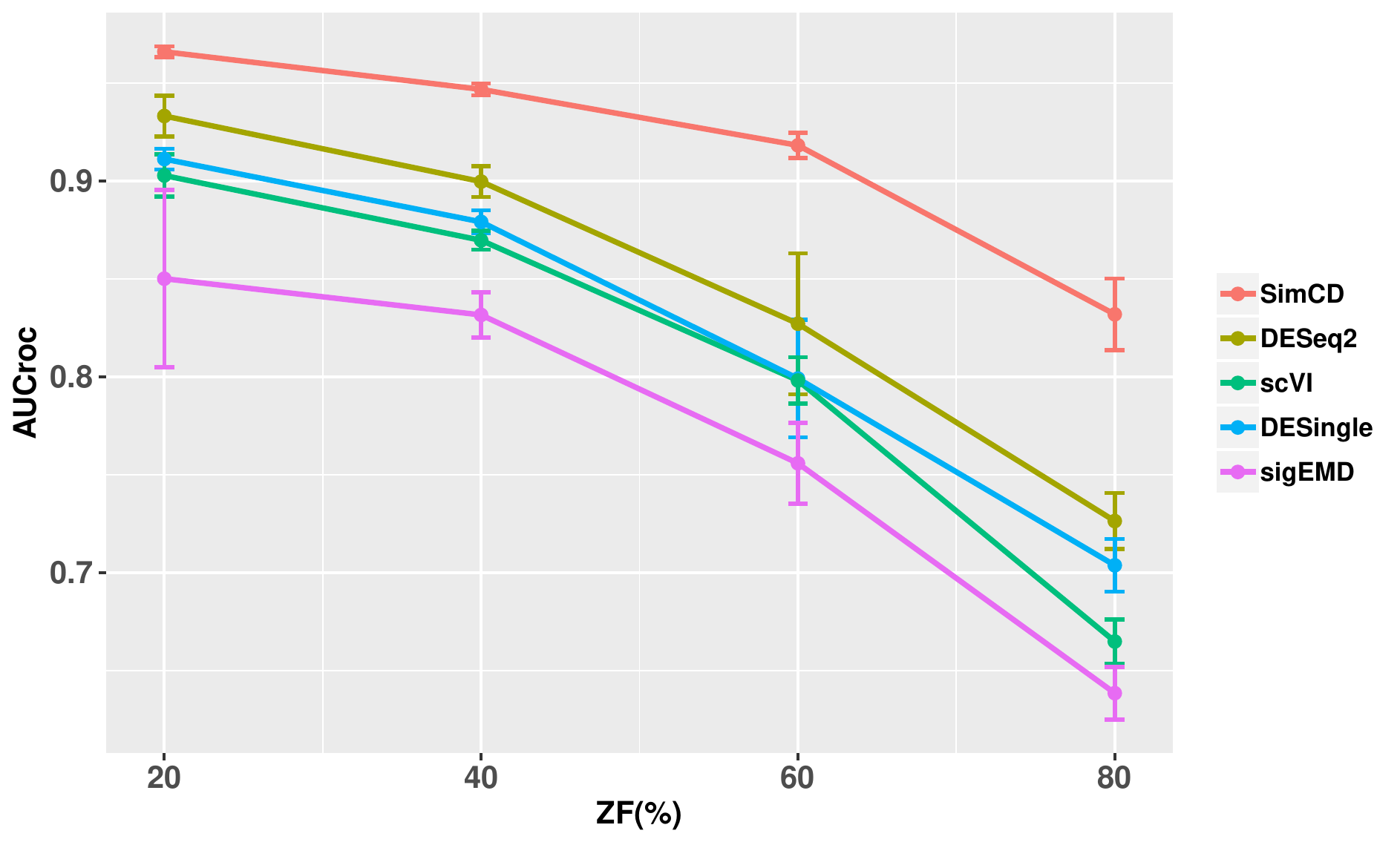}
    \end{subfigure}\hspace{2mm}
    \begin{subfigure}{.45\textwidth}
        \includegraphics[width=\textwidth,keepaspectratio,trim=5 5 90 10, clip]{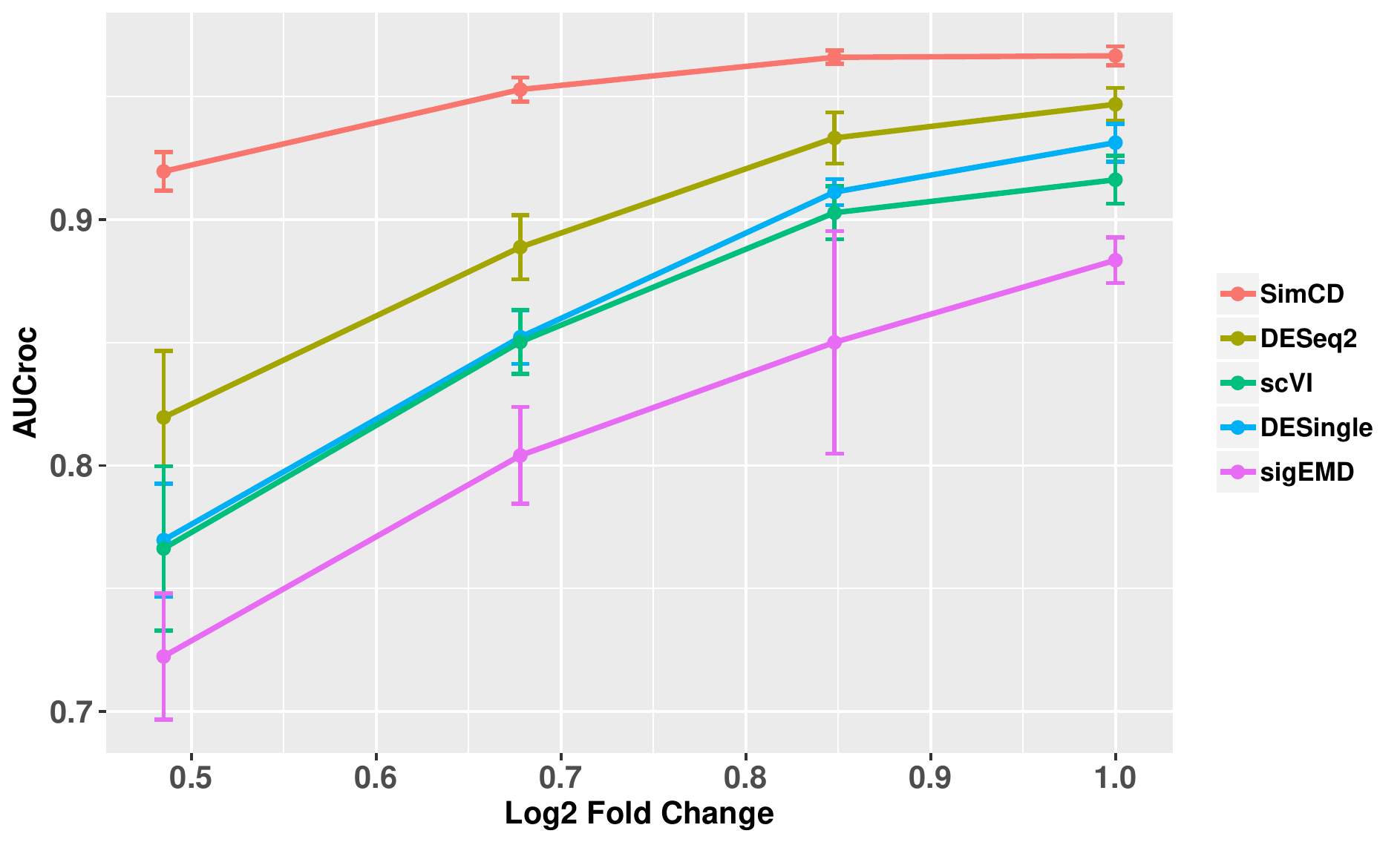}
    \end{subfigure}\vspace{-2mm}
    \caption{AUC-ROC of DE analyses with hGNB simulated data for different zero fractions (left) and log2 fold changes (right).}\vspace{-5mm}
    \label{fig:de}
\end{figure*}

In the first simulation study, we generate the synthetic scRNA-seq data for 10000 genes under two conditions according to the hGNB generative model. Each of the conditions has fifty replicates, i.e. 100 cell samples in total. To make the synthetic data closely resemble real-world scRNA-seq data, the parameters of hGNB are first inferred from the mouse hypothalamus neuron cell types scRNA-seq dataset~\citep{mousehypo2017} and then synthetic sequencing counts are generated using these inferred model parameters. In this simulation setup, the expression of gene $g$ in sample $j$ is simulated from $ \mbox{NB}(r_j, p_{gj})$, where $p_{gj}$ follows \eqref{eq:simcd}. For $j\in\{1,2,...,J\}$, the design matrix elements are
$\bold{x}_{j}^{(1)}$ = [$x_{0j}^{(1)}$ , $x_{1j}^{(1)}$] and $\bold{x}_{g}^{(2)}$ = [$x_{0g}^{(2)}$]. In this simulation setting, $m=n=0$ corresponds to the intercept term in gene- and cell(sample)-relevant covariates and the element corresponding to $n=1$ corresponds to the cell condition. More precisely, $x_{1j}^{(1)}$ = 0 if no treatment has been applied to sample $j$, and $x_{1j}^{(1)}$ = 1 if this sample is under treatment.

The effect of the covariate corresponding to the $n$th design matrix element on the expression level of gene $g$ is adjusted through the regression coefficient $\beta_{gn}^{(1)}$. We simulate this coefficient according to a zero-mean normal distribution with precision parameter $\alpha_n$. For the condition covariate, we draw the precision parameter as $\alpha_1 \sim\mbox{Gamma}(7.87e4,1/1e4)$. Under this setting, with 10\% probability, the absolute value of $\beta_{g1}^{(1)}$ is larger than 0.588. Thus on average, 10\% of genes exhibit an expression fold change of at least $\exp(0.588) = 1.8$ between the two different conditions. To simulate cell clusters, a $K$(=3, number of latent factors)-variate normal mixture distribution with three components is fitted to the inferred score parameter $\theta_j$ from real-world data and then for each simulated dataset, scores are generated from the K-variate normal distributions. By adjusting the mean parameters of the normal distributions, from which we sample $x_{mg}^{(2)}$ values, we generate synthetic datasets at four levels of zero-count fractions as 20\%, 40\%, 60\% and 80\%. For each zero-count percentage level, we simulate 10 independent datasets. Then, we benchmark the performance of SimCD in terms of the differential expression analysis with those of DESeq2~\citep{deseq2}, DESingle~\citep{DEsingle}, and sigEMD~\citep{sigEMD}, which are three popular DE methods. We also compare with scVI on the differential expression analysis performance. For SimCD, model parameters are inferred via Gibbs sampling, where in each run of the algorithm, we collect 1000 MCMC samples after 1000 burn-in iterations and then rank the genes by the symmetric KL-divergence measure developed in Section 2.3. For DESingle, sigEMD and scVI we follow their default analysis pipelines. DESeq2 was originally designed to perform DE analysis on bulk RNA-seq data. In order to perform DE analysis on scRNA-seq data using DESeq2, we follow the recommendations in~\cite{vandenberge2018} to use the phyloseq normalization and LRT test over Wald test. Figure~\ref{fig:de}(left) presents the area under ROC curves (AUC-ROC) of SimCD, DESeq2, DESingle, scVI and sigEMD at four different zero fraction~(ZF) levels. SimCD clearly outperforms other methods at all four levels in terms of AUC-ROC. Furthermore, we investigate the robustness of our DE results to different simulated log2 fold change values in Figure~\ref{fig:de}(right), which shows that SimCD performs better than other methods with a significant margin. This validates the benefit of accounting for cell heterogeneity when doing DE analysis to better capture the dynamic expression changes.

We also benchmark the cell clustering performance of SimCD with those of scVI and ZINB-WaVE in this simulation setup, as these are either the most similar model as SimCD or have been reported with the SOTA clustering results. We evaluate the clustering performance based on the average silhouette width~(ASW) measure. The silhouette width $s(j)$ for sample $j$ is defined as
$s(j) = {\textstyle \frac{b(j) - a(j)}{max\{a(j),b(j)\}}}$,
where $a(j)$ is the average within-distance of sample $j$  and $b(j)$ is the minimum average distance between sample $j$ and samples in other clusters. 
Figure~\ref{fig:clustering}(left) provides the ASW values of the clustering results by SimCD, ZINB-WaVE and scVI at four different zero-count percentages. As the figure suggests, SimCD has the highest ASW for all cases. These results illustrate the capability of SimCD to infer the cell heterogeneity, even at high zero-count prevalence, by simultaneous modeling of gene-level expression changes and proper hierarchical structure. 

All the results are also provided in the tables in Appendix A.3. 

\subsubsection{Comparison on data simulated by the ZINB-WaVE model}

\begin{figure*}[t!]
    \centering
    \begin{subfigure}{.49\textwidth}
        \includegraphics[width=\textwidth,keepaspectratio, trim=5 5 10 10, clip ]{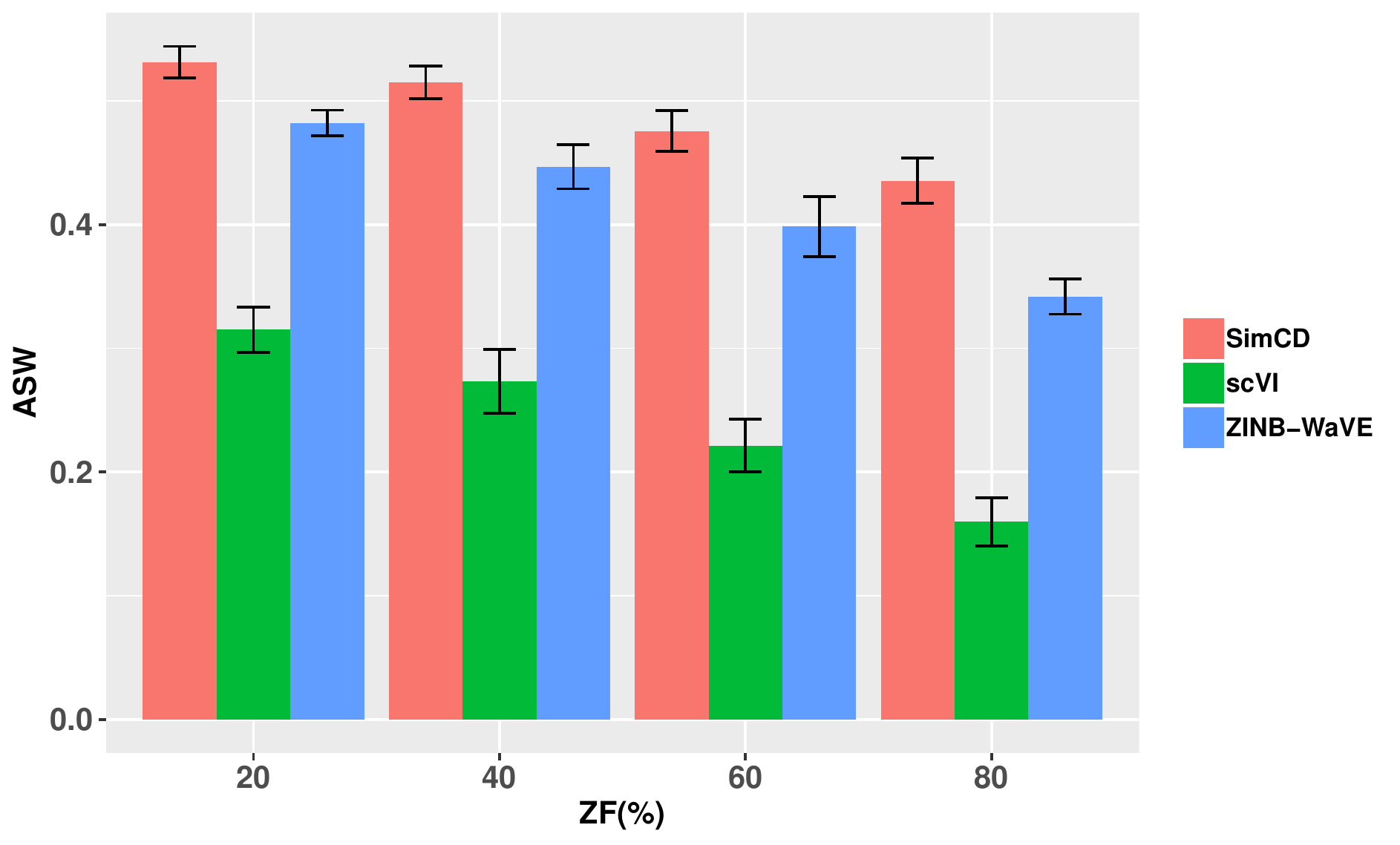}
    \end{subfigure}\hspace{0.8mm}
    \begin{subfigure}{.5\textwidth}
        \includegraphics[width=\textwidth,keepaspectratio,trim=5 5 10 5, clip]{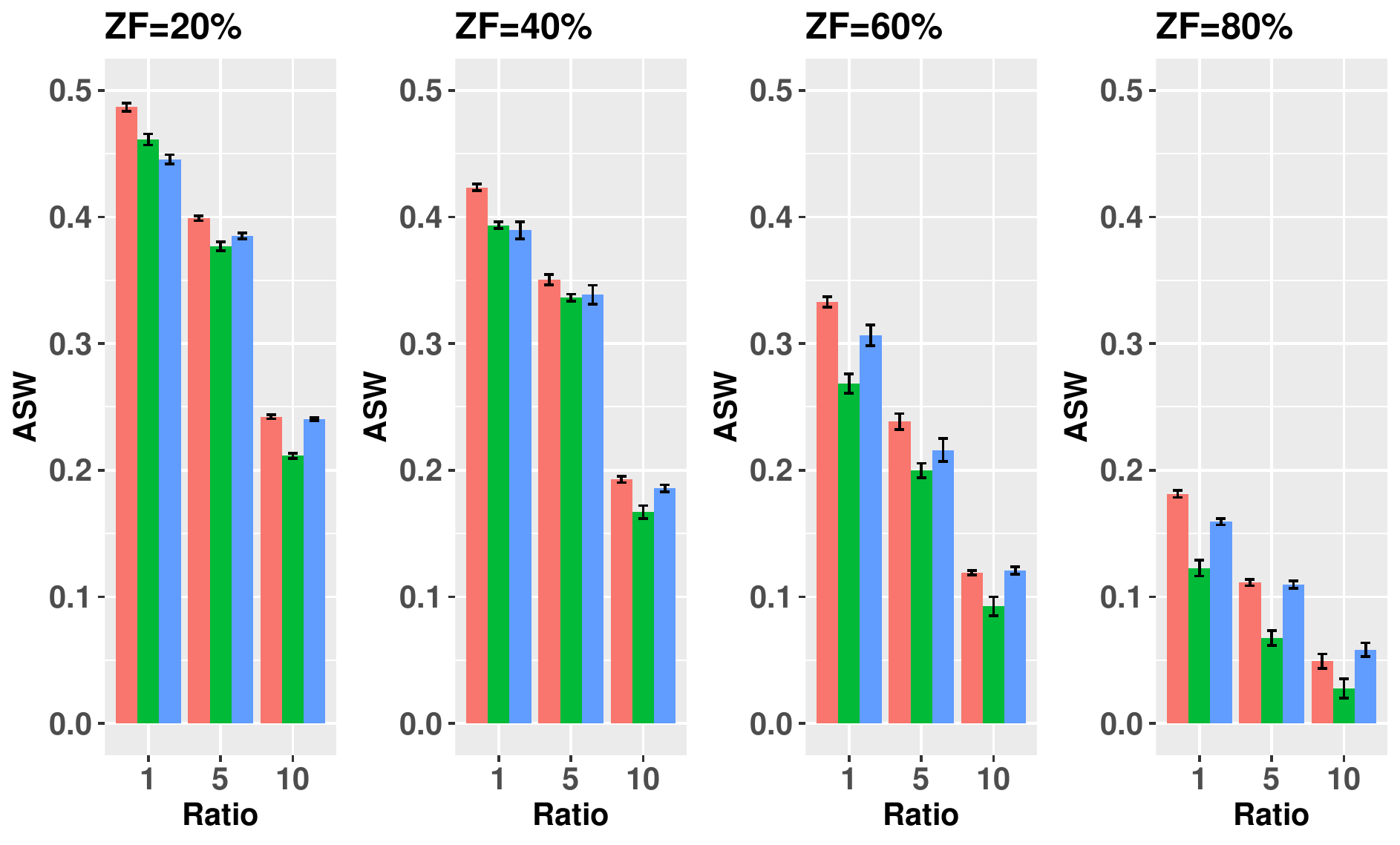}
    \end{subfigure}\vspace{-2mm}
    \caption{Clustering ASW with hGNB (left) and ZINB-WaVE (right) simulated data for different zero fractions.}
    \label{fig:clustering}
\end{figure*}

\begin{table*}[!b]
\centering
\caption{{AUC-ROC of DE analyses with the ZINB-WaVE simulated data for different zero fraction (ZF) levels (RZ: Real Zeros, IZ: Inflation Zeros). }
\label{Tab:deSINB}}
\resizebox{0.75\textwidth}{!}{
{\begin{tabular}{@{}c|c|c|ccccc@{}}
\toprule {\bf ZF} & {\bf RZ} & {\bf IZ} & {\bf SimCD} & {\bf DESeq2} & {\bf DESingle} & {\bf scVI} & {\bf sigEMD}\\\hline \hline
{} & {15\%} & 5\% & {\bf 0.9614} $\pm$ 0.0036 & 0.9164 $\pm$ 0.0054 & 0.9254 $\pm$ 0.0102 & 0.8780 $\pm$ 0.0065 & 0.8812 $\pm$ 0.0113 \\
{\bf 20\%} & {10\%} & 10\%  & {\bf 0.9592} $\pm$ 0.0054 & 0.8931 $\pm$ 0.0084 & 0.9145 $\pm$ 0.0148 & 0.8737 $\pm$ 0.0064 & 0.8354 $\pm$ 0.0088\\
{} & {5\%} & 15\% & {\bf 0.9482} $\pm$ 0.0103 & 0.8407 $\pm$ 0.0014 & 0.8950 $\pm$ 0.0102 & 0.8364 $\pm$ 0.0169 & 0.8499 $\pm$ 0.0150 \\ \hline
{} & {30\%} &  10\% & {\bf 0.9339} $\pm$ 0.0085 & 0.8563 $\pm$ 0.0077 & 0.8385 $\pm$ 0.0198 & 0.8387 $\pm$ 0.0054 & 0.8233 $\pm$ 0.0144 \\
{\bf 40\%} & {20\%} & 20\%  & {\bf 0.9069} $\pm$ 0.0164 & 0.8155 $\pm$ 0.0213 & 0.8504 $\pm$ 0.0105 & 0.7906 $\pm$ 0.0095 & 0.8273 $\pm$ 0.0078\\
{} & {10\%} & {30\%} & {\bf 0.8884}  $\pm$ 0.0024 &  0.7861 $\pm$ 0.0211 &  0.8753 $\pm$ 0.0124 &  0.7606 $\pm$ 0.0110 &  0.8300 $\pm$ 0.0256 \\ \hline
{} & {45\%} & 15\% & {\bf 0.8336} $\pm$ 0.0082 & 0.7511 $\pm$ 0.0276 & 0.7335 $\pm$ 0.0185 & 0.6260 $\pm$ 0.0312 & 0.7317 $\pm$ 0.0221 \\
{\bf 60\%} & {30\%} & 30\%  & {\bf 0.7795} $\pm$ 0.0091 & 0.6838 $\pm$ 0.0125 & 0.7656 $\pm$ 0.0123 & 0.6193 $\pm$ 0.0065 & 0.7556 $\pm$ 0.0105\\
{} & {15\%} & 45\% & 0.7156 $\pm$ 0.0106 & 0.6377 $\pm$ 0.0214 & {\bf 0.8227} $\pm$ 0.0164 & 0.5545 $\pm$ 0.0098 & 0.7752 $\pm$ 0.0207 \\ \hline
{} & {60\%} &  20\% & {\bf 0.6503} $\pm$ 0.0125 & 0.5878 $\pm$ 0.0102 & 0.5958 $\pm$ 0.0164 & 0.5203 $\pm$ 0.0085 & 0.6184 $\pm$ 0.0260 \\
{\bf 80\%} & {40\%} & 40\% & 0.5667 $\pm$ 0.0085 & 0.5476 $\pm$ 0.0057 & {\bf 0.6273} $\pm$ 0.0134 & 0.5123 $\pm$ 0.0154 & 0.6256 $\pm$ 0.0191\\
{} & {20\%} & {60\%} & 0.5434 $\pm$ 0.0194 &  0.5381 $\pm$ 0.0267 & {\bf 0.7061} $\pm$ 0.0278 &  0.5051 $\pm$ 0.0314 &  0.6821 $\pm$ 0.0357 \\ \hline
\end{tabular}}{}
}
\end{table*}

In the second simulation study, we simulate synthetic scRNA-seq datasets from the ZINB-WaVE model~\citep{zinbwave2018} based on the ZINB distribution. By employing this generative model, which is different from the underlying hGNB model for SimCD, we study the robustness of SimCD to the model mismatch as well as varying levels of zero inflation simulated by varying the parameters of the ZINB distribution. We also infer the ZINB-WaVE model parameters from the scRNA-seq dataset of mouse hypothalamus neuron cell types~\citep{mousehypo2017} that has both dynamic gene expression changes and cell-level heterogeneity. Then we generate count data for 10000 genes in 100 samples across two conditions. By adjusting the values of regression coefficients in the ZINB-WaVE model, we generate synthetic datasets with four levels of zero-count percentage as 20\%, 40\%, 60\% and 80\%. For each of the four zero fractions, we considered three different ratios of zero counts directly coming from the NB distribution component (as ``real zeros'') to zeros from the zero inflation term in the ZINB model (``inflation zeros''), leading to the total of 12 cases. For each case we generate 5 independent datasets. Moreover, for each simulated dataset we change the regression coefficients in the ZINB distribution mean parameter term so that on average 10\% of genes show differential gene expression patterns across two conditions. We also simulate cell clustering structures with different ratios of within- to between-cluster sums of squared distances~(SSD) in the ZINB-WaVE model (details can be found in Appendix A.2).

Table~\ref{Tab:deSINB} provides the AUC-ROC of SimCD, DESeq2, DESingle, scVI and sigEMD at twelve different zero fraction setups with inflation (technical) and real~(biological) zero-count percentages. As the table suggests, SimCD outperforms the competing methods with a significant margin for all cases except those having more than 40\% of counts simulated from the inflation term in the ZINB-WaVE model. In these cases, DESingle outperforms other methods. This is reasonable as DESingle is based on the ZINB regression model, which can better estimate the proportion of the real and inflation zeros in this set of simulated data~\citep{DEsingle}. However, we should note that SimCD still performs better than scVI and DESeq2 in these scenarios, suggesting the robustness of our SimCD to identify the differentially expressed genes even with the mismatched simulation model in presence of high inflated zero percentages without explicit inflation modeling.

We next compare SimCD with ZINB-WaVE and scVI in terms of their ability to detect the cell clustering structures in the simulated scRNA-seq datasets. ASW values for the clustering results are shown in Figure~\ref{fig:clustering}(right). Again, SimCD detects simulated clusters better than scVI and ZINB-WaVE, except for the cases where the simulated ratio of within- to between-cluster SSD is 10 (harder clustering problem) and the zero fraction is either 60\% or 80\%. In these situations ZINB-WaVE performs better than SimCD as the data are generated from its own model. Excessive number of inflation zeros and complexity of cell structure in the simulated count data in these cases lead to the degraded performance of SimCD. However, SimCD is still doing better than scVI and moreover, its performance is comparable with ZINB-WaVE. 

Overall, DE and clustering analyses of the ZINB-WaVE simulated count data clearly show the advantage of performing these two tasks simultaneously by SimCD over doing it separately using the methods that are optimized for one task like DESingle, sigEMD, DESeq2, and ZINB-WaVE or using already developed methods that can do DE analysis and clustering together, such as scVI.

\subsection{Comparison on real-world scRNA-seq data}

\begin{figure*}[t!]
    \centering
        \includegraphics[width=0.9\textwidth,keepaspectratio]{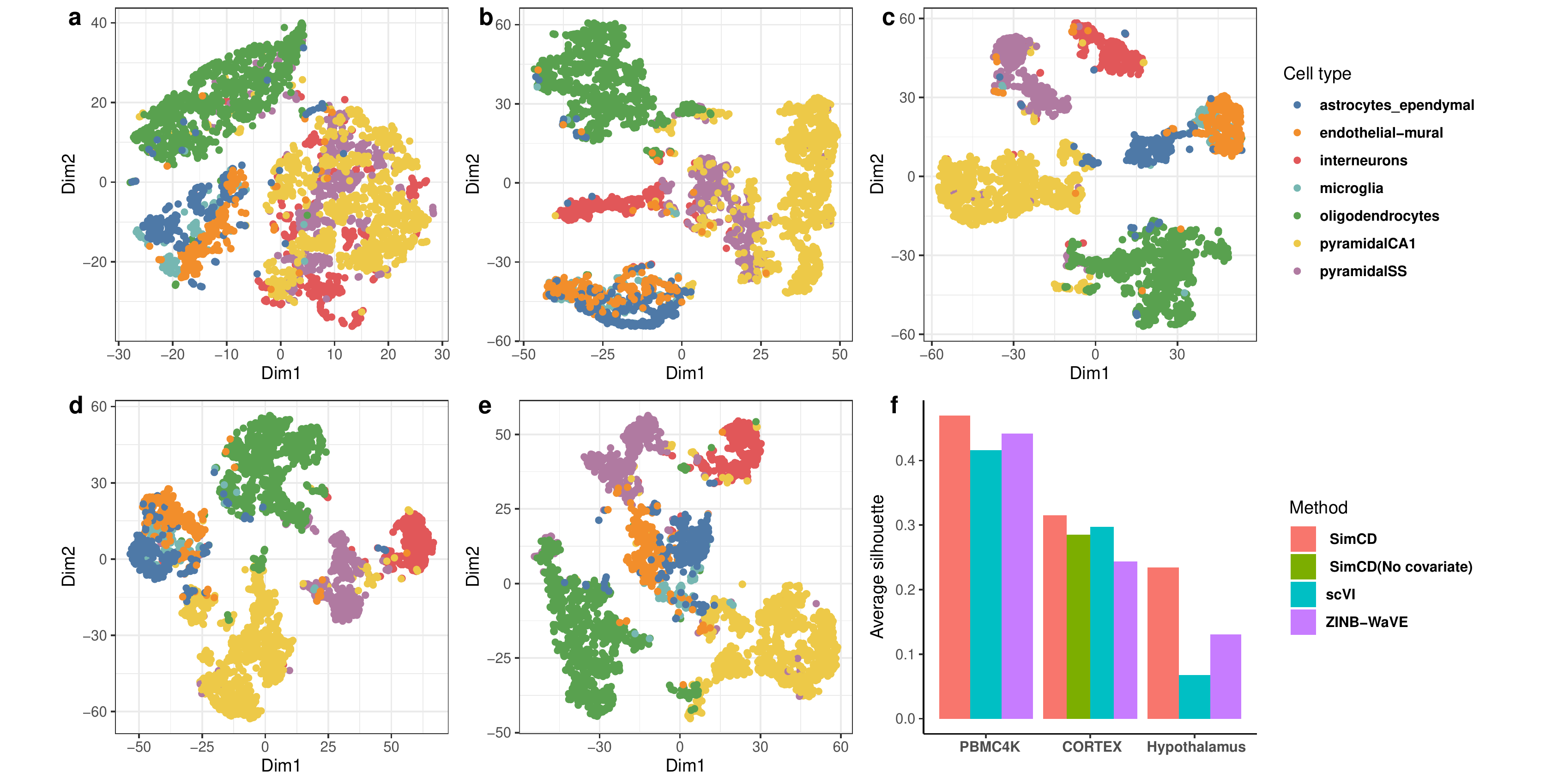}\vspace{-3mm}
        \caption{Two-dimensional t-SNE visualization based on the (a) raw CORTEX dataset, and derived latent representations by (b) ZINB-WaVE, (c) scVI, (d) SimCD (no covariate), and (e) SimCD. The bar plots in (f) illustrate the comparison of the clustering results by ASW for PBMC4k, CORTEX and mouse hypothalamus datasets.}
        \label{Fig:multtsne}
\end{figure*}
We further benchmark SimCD with other SOTAs on real-world scRNA-seq datasets with respect to both clustering and DE results when the ground-truth annotations are available. 

\subsubsection{CORTEX dataset}

This dataset characterizes 3005 mouse cortex cells using Fluidigm C1 microfluidics cell capture platform followed by Illumina sequencing \citep{Zeisel2015}. Single-cell gene expression is quantified by UMI counts. In addition to gene expression data, additional annotations of the samples, including cell cluster labels, age and sex of the corresponding mice, are also available in this dataset. We retain the top 558 genes ordered by variance for analyses following \cite{lopez2018deep}. We compare the cell clustering performance of SimCD with those of scVI and ZINB-WaVE on this dataset. When applying SimCD on the CORTEX dataset we consider two different setups. In the first setup, we use the available age and sex metadata as confounding cell covariates when designing $x_{nj}^{(1)}$ in the model \eqref{eq:simcd}. More specifically, $x_{1j}^{(1)}$ is 1 or 0 based on sex of sample $j$ (male or female). Age in metadata is normalized within $(0,1)$ and used as $x_{2j}^{(1)}$. In the second setup, we do not use age and sex as cell covariates and $x_{nj}^{(1)}$ is just the fixed intercept term $x_{0j}^{(1)}$. The second setup is labelled ``SimCD (no covariate)'' in the following results. For scVI and ZINB-Wave, we follow their default parameter and pipeline setups as indicated in their original publications. 
The run-time of SimCD with 2000 MCMC sampling iterations on the cluster node with configuration provided earlier in Section 3 is around 6 hours.

Figures \ref{Fig:multtsne}a-e show the two dimensional visualization of the CORTEX count data by t-Stochastic Neighborhood Embedding~(t-SNE)~\citep{Hinton_Roweis_2003} based on the raw data and the derived latent representations by ZINB-WaVE, scVI, SimCD with no covariate, and SimCD, respectively. The corresponding cells are colored by the given cell cluster label annotations in the dataset. As shown in the figure, SimCD and scVI distinguish pyramidal SS and pyramidal CA1 or enothelial-mural and astrocytes\_ependymal cell clusters while ZINB-WaVE fails to accomplish these tasks. The derived latent representations are used to perform cell clustering by {\tt RSEC} implemented in the Bioconductor package {\tt clusterExperiment}~\citep{cluster2017}. 
We calculate ASW values for the clustering results by each method to benchmark their cell clustering performances. As shown in Figure \ref{Fig:multtsne}f, SimCD outperforms scVI and ZINB-WaVE with higher ASW values. Additionally, SimCD's clustering performance can be further improved when adjusting for potential confounding effects by incorporating age and sex as cell-level covariates, comparing to SimCD without these covariates in the model (more details can be found in Appendix Figure S1). This is expected as accounting for known biological factors can help better model the cell clustering structures in the scRNA-seq data by disentangling them from the main biological factor(s) affecting single-cell gene expression changes, in which we are interested.

\subsubsection{PBMC dataset}

\begin{figure*}[t!]
    \centering
        \includegraphics[width=\textwidth,keepaspectratio]{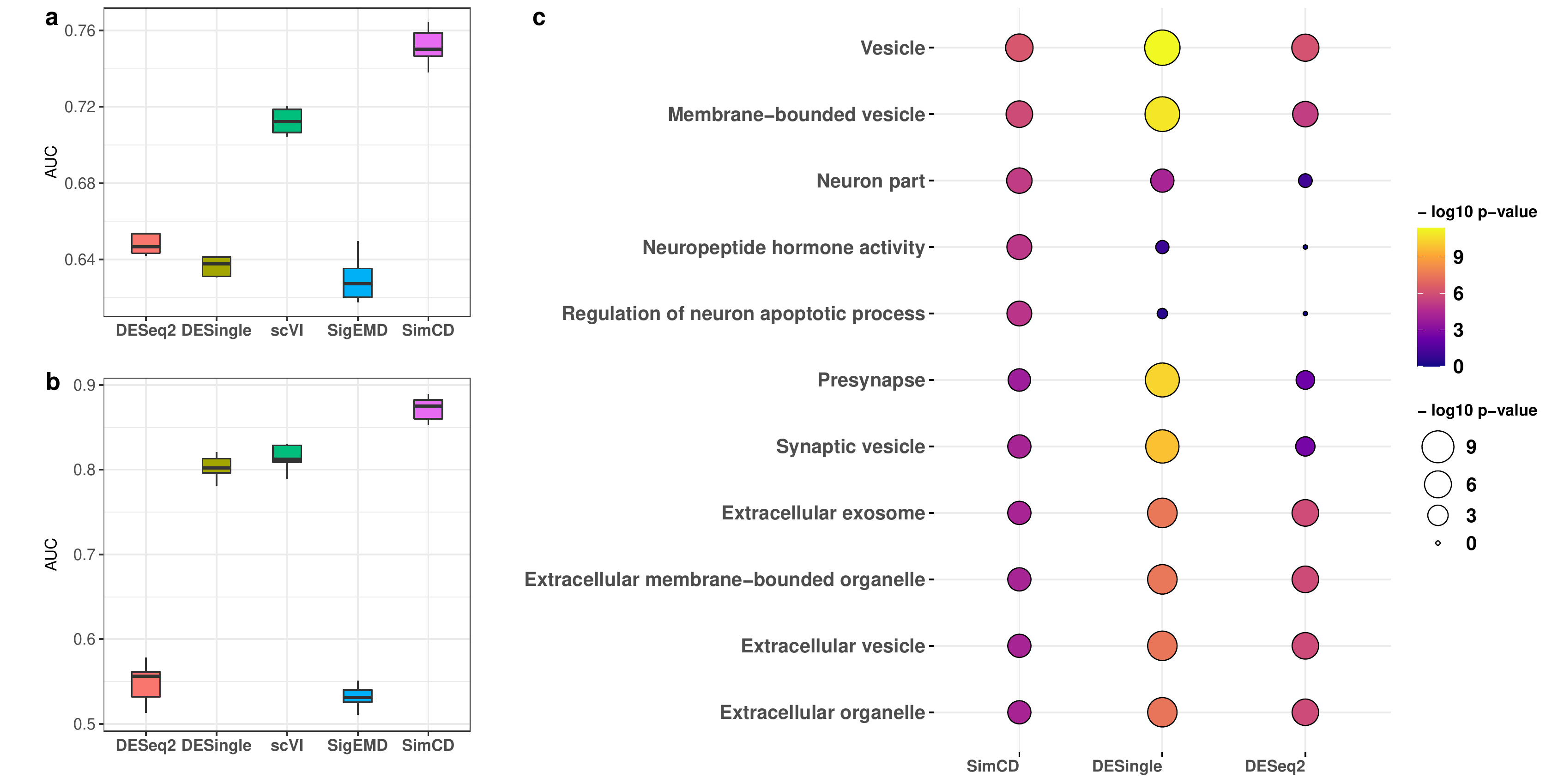}\vspace{-3mm}
        \caption{DE performance evaluation based on the consistency with the pseudo ground-truth extracted from bulk data: (a) between B and dendritic cells, and (b) between CD$4^+$ and CD$8^+$ cells from PBMC data. GO enrichment analysis comparison of different methods in (c) shows -log10 p-values for the union set of top five enriched GO terms of top DE genes from SimCD, DESingle and DESeq2 in the mouse hypothalamic case study.}
        \label{Fig:multDE}
\end{figure*}

We further investigate the performance of SimCD on another scRNA-seq dataset from two batches of peripheral blood mononuclear cells (PBMC) from a healthy donor (PBMC4k and PBMC8k)~\citep{pbmc2017}, which has been analyzed by multiple scRNA-seq analysis methods. After pre-processing and filtering as in \cite{lopez2018deep} and \cite{scone2017}, 12039 cells and 10319 genes were retained. We first focus on the cells from the PBMC4k batch and 1000 genes with the highest variances across the cells from this batch to assess the clustering performance of SimCD. We use the derived cell cluster labels by {\tt Seurat} as the ground truth labels, as the authors in \cite{lopez2018deep} have validated its biological significance. Figure \ref{Fig:multtsne}f again demonstrates the superior performance of SimCD in terms of having a higher ASW value and thus more biologically meaningful cell clustering. Figures \ref{Fig:multtsne_hypo_pbmc}a-d provide the t-SNE visualization of the derived latent representations by different competing methods on this dataset, showing tighter clusters consistent with the ground-truth clustering labels.    

Next, we evaluate the differential expression (DE) performance of SimCD by performing DE analysis between B cell and dendritic cell clusters and between CD$4^+$ and CD$8^+$ T cell clusters in 12309 filtered cells combining the PBMC4k and PBMC8k batches. Similar to \cite{lopez2018deep}, bulk microarray-based results between the mentioned cell groups served as the ground truth. Specifically, genes that have adjusted p-values below 0.05 in the bulk RNA-seq analysis are considered as true differentially expressed genes. We filter out genes that we can not find in bulk data DE analysis, leading to remaining 3346 genes. In order to have more robust results we randomly sample 200 cells from each cluster 10 times and calculate AUCROC for each set. Figures \ref{Fig:multDE}a and b show the box plots of AUCROC values for 10 independent runs of DE analysis by different methods between B and dendritic cell clusters, as well as between CD$4^+$ and CD$8^+$ T cell clusters. Figure \ref{Fig:multDE}a indicates that on average SimCD achieves the highest AUC value followed by scVI and then DESeq2 when comparing B and dendritic cell clusters. Additionally, results of comparing CD$4^+$ and CD$8^+$ cell clusters illustrate that again SimCD clearly performs better than other methods in terms of having more consistent DE analysis results with the DE results based on the bulk data analysis as presented in Figure \ref{Fig:multDE}b. 

All these results are also provided in the tables in Appendix A.4.



\subsection{Mouse hypothalamus case study}

\begin{figure*}[!t]
    \centering 
        \includegraphics[width=\textwidth,keepaspectratio]{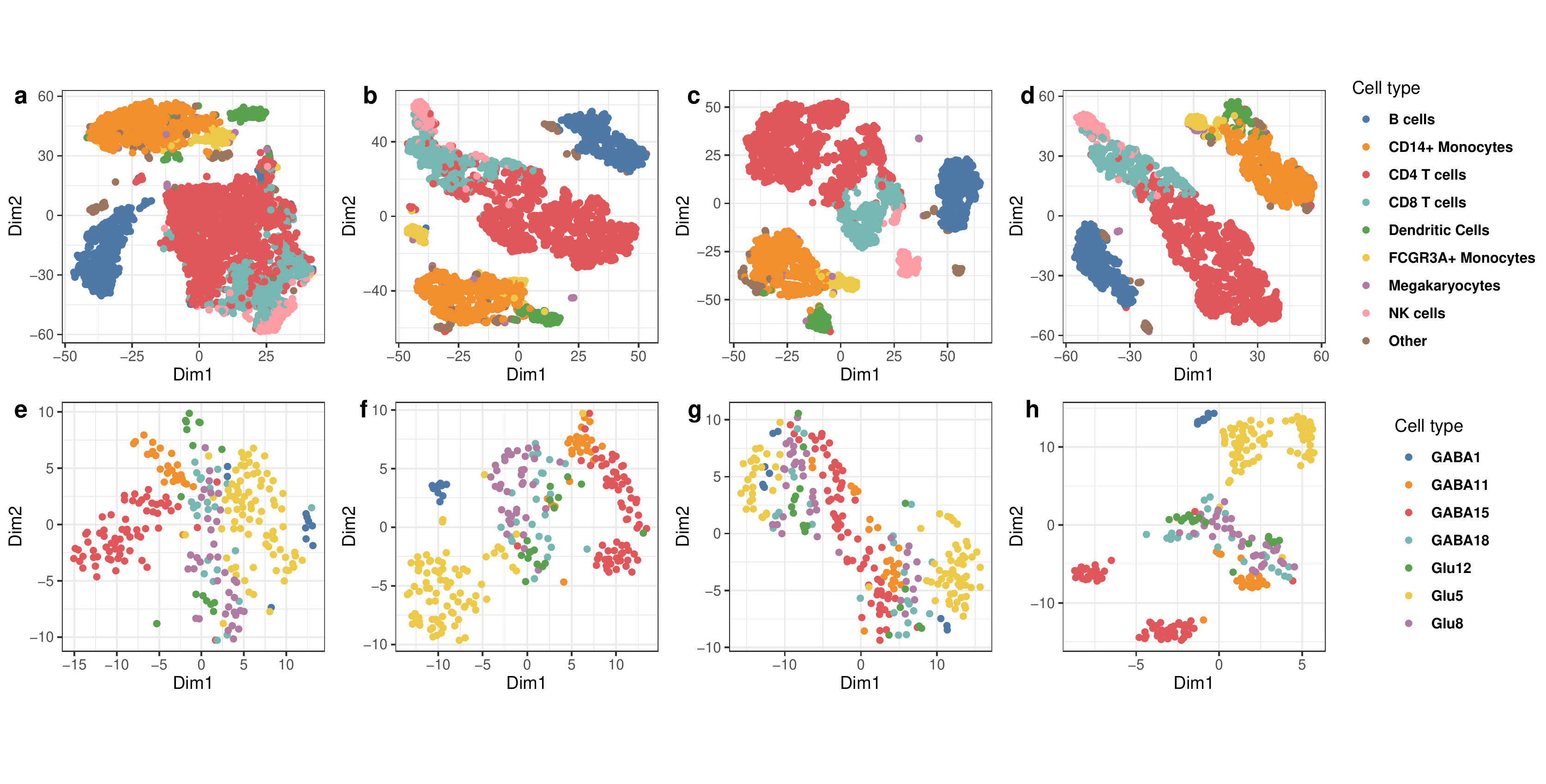}\vspace{-1mm}
        \caption{\textbf{Top:} Two-dimensional t-SNE visualization of the (a) raw data and the derived latent representations by (b) ZINB-WaVE, (c) scVI, and (d) SimCD for the PBMC4k dataset. \textbf{Bottom:} Two-dimensional t-SNE visualization of the (e) raw count data and derived latent representations by (f) ZINB-WaVE, (g) scVI, and (h) SimCD for the mouse hypothalamus dataset. }
        \label{Fig:multtsne_hypo_pbmc}
\end{figure*}

We further illustrate the utility of SimCD to reveal the underlying biological signals inherent in scRNA-seq data in a case study on analyzing mouse hypothalamus scRNA-seq count data. Hypothalamus is an important brain region regulating important cellular processes such as feeding and metabolism. In particular, understanding the cell composition and thereafter identifying the neuronal cell subtypes in hypothalamus helps to gain insights into the inherent biological modules involved in regulating feeding behavior. To this end, we use the scRNA-seq dataset in~\cite{mousehypo2017} with gene expression profiles of 45 cell clusters, including 34 neuronal and 11 non-neuronal cell subtypes. Sequencing the hypothalamus cells in four normally-fed and three food-deprived mice makes this dataset suitable to evaluate the transcriptional response of mouse hypothalamus cells to food-deprivation. The accession number of this dataset is GSE87544. Here, we focus on seven neuronal clusters Glu5, Glu8, Glu12, GABA1, GABA11, GABA15 and GABA18 reported in~\cite{mousehypo2017} showing differential gene expression response to food deprivation. After filtering out genes that have count per million~(CPM) below 1 (low expressed genes), we have scRNA-seq count data of 12850 genes in 263 cells, of which 104 cells are from  normally-fed mice and the rest are food-deprived (Appendix Table S8). 

We apply SimCD to this dataset to simultaneously identify differentially expressed genes between cells from hungry and normal mice in the above seven cell clusters and perform cell clustering. For $j\in\{1,2,...,J\}$, the design matrix elements are 
$\bold{x}_{j}^{(1)}$ = [$x_{0j}^{(1)}$, $x_{1j}^{(1)}$] and $\bold{x}_{g}^{(2)}$ = [$x_{0g}^{(2)}$]. In this case study, $x_{0j}^{(1)} = x_{0g}^{(2)} = 1$ corresponds to the intercept term in gene- and cell(sample)-relevant covariates and the element corresponding to $n=1$ corresponds to the cell condition. More precisely, $x_{1j}^{(1)}=0$ if cell $j$ is from normal mice, and $x_{1j}^{(1)}=1$ if it is from hungry mice. During the inference process, we collect 1000 MCMC samples after 1000 burn-in iterations to calculate the symmetric KL-divergence and infer the cell clusters in the dataset. 

Figure \ref{Fig:multtsne}f shows that SimCD clearly outperforms scVI and ZINB-WaVE as it generates more biologically meaningful cell clusters based on the annotated seven neuronal cell clusters. Poor clustering performance of scVI on this dataset can be due to the fact that the number of cells is much smaller than the number of genes in this dataset, which can result in the inductive bias of the deep neural networks implemented in scVI~\citep{lopez2018deep}. Figures \ref{Fig:multtsne_hypo_pbmc}e-h visualize the t-SNE plots of the latent representations derived by all competing methods.

In terms of DE analysis results, When checking the ten most differentially expressed genes based on their KL values calculated by SimCD, we find that most of them have been previously reported as significant modules involving in feeding process. The top differentially expressed gene detected by SimCD is Galanin~(\emph{Gal}), a neuropeptide, whose role in regulation of appetite, food behavior and food reward has been previously reported in \citep{Gal2003,Gal1986,Gal2017}. The second gene is hypocretin (also known as Orexin; shown by \emph{Hcrt}), another neuropeptide that regulates metabolism, appetite and arousal \citep{LHAhypo2019}. The third gene, Growth hormone-releasing hormone (\emph{Ghrh}), has been shown to play the key role in metabolism and is also responsive to food-deprivation \citep{Ghrh1997,Ghrh1993}. Neurotensin~(\emph{Nts}) is ranked as the fourth most differentially expressed gene by SimCD. This Neuropeptide has been shown recently to be involved in feeding process and weight loss behaviors \citep{Nts2018,Nts2017}. The rest of genes in the list are \emph{Ttc3}, \emph{Synpr}, \emph{Npepps}, \emph{Nrxn1}, \emph{Cirbp} and \emph{Luc7l3}. \emph{Cirbp} has also been confirmed by immunostaining to have increased expression levels in Glu5 (MM neurons) upon food-deprivation \citep{mousehypo2017}. The other top differentially expressed genes detected by SimCD can be potential bio-markers for the response to food deprivation involved in feeding process. 

To further demonstrate the biological significance of the detected genes, we consider the genes having KL values greater than 2 by SimCD as differentially expressed ones, leading to 107 DE genes. We also perform DE analysis between normally-fed and food-deprived cells by DESingle and DESeq2 as well, lead to the corresponding lists of top 100 genes that have lowest adjusted p-values. We exclude scVI from this analysis since based on the clustering results by scVI, it cannot fit this dataset well because of the small sample size compared to the number of genes. We also exclude SigEMD due to its overall inferior DE performance on previous simulated and real-world scRNA-seq data. We then perform Gene Ontology (GO) analysis of the top DE genes from each method, covering molecular function (MF), cellular component (CC) and biological process (BP) ontology domains. The top ten significantly enriched GO terms with their corresponding P-values for SimCD, DESingle and DESeq2 are shown in Appendix A.5. We retain the top five enriched GO terms from each of these three methods, leading to a union set of 11 unique GO terms (for example, GO term \emph{Vesicle} is the common top GO term by all three methods; \emph{Membrane-bounded vesicle} is the second top enriched GO term by both SimCD and DESingle). Figure \ref{Fig:multDE}c shows the negative log10 p-values of those 11 highly enriched GO terms from SimCD, DESingle, and DESeq2, by GO enrichment analysis. Note that the top five GO terms enriched in DESignle and DESeq2 are also highly enriched in SimCD DE analysis results as well. On the other hand, the GO term \emph{Regulation of neuron apoptotic process}, which has been previously reported in the GO analysis of the top DE genes between \emph{ad lib} fed and food-deprived mice by bulk mRNA analysis~\citep{Jiang2015}, is only highly enriched in the top DE genes detected by SimCD. 

Moreover, enriched GO terms based on the identified DE genes by SimCD agree with the current biological understanding of the response to food deprivation. Presence of GO terms, such as \emph{Neuron part}, \emph{neuronal cell body}, and \emph{Neuropeptide hormone activity} in the top ten enriched GO terms by SimCD, confirms the neuron identity and differentiation inherent in the scRNA-seq data in this case study~\citep{mousehypo2017}. These results illustrate the association of the identified DE genes by SimCD with neuronal dynamic processes in hypothalamic neuronal subtypes responding to food deprivation. 

Overall, by comparing clustering and DE analysis results by SimCD in this case study with those of SOTAs specially developed for one of these two tasks, it is clear that having a unified model that can learn both clustering structures and dynamic differential expression changes inherent in scRNA-seq data can provide additional meaningful biological insight that cannot be identified by other separately designed tools for either DE analysis or cell clustering.     

\section{Conclusions}

SimCD has been proposed to simultaneously cluster cells and detect differentially expressed genes for scRNA-seq count data. SimCD is based on a unified hGNB model that inherently models cell heterogeneity and account for both gene- and cell-level biological factors to better model zero-inflation (sparsity) in scRNA-seq data. SimCD obviates the need for any pre-processing step, such as normalization, thanks to explicit modeling of different covariate effects. By providing extensive results on simulated and real-world scRNA-seq data, we demonstrate that SimCD can outperform the state-of-the-art methods for scRNA-seq clustering and differential expression analysis, which have been often developed disjointly for these two tasks. Future work concerns further improving the efficiency of the inference algorithm as well as the model expressive power by introducing semi-implicit variational distributions~\citep{Shahin} to SimCD.

%
%






\title{SimCD: Simultaneous Clustering and Differential expression analysis for single-cell transcriptomic data: Supplementary Materials}
\maketitle


\clearpage
\section*{A. Appendix}

\subsection*{A.1 Gibbs sampling inference}

We provide the detailed Gibbs sampling procedure by exploiting 
the augmentation techniques for the negative binomial (NB) distribution~\citep{zhou2015negative} and the Polya-Gamma (PG) distributed auxiliary variable technique~\citep{LGNB_ICML2012,polson2013bayesian}.

\noindent {\bf Sampling $r_j$.\quad}
By exploiting the  data augmentation techniques in ~\cite{zhou2015negative}, we implement an efficient Gibbs sampling algorithm with closed-form updating steps. More precisely, we infer the cell-dependent dispersion parameter of the NB distribution by first drawing latent random counts from the Chinese Restaurant Table~(CRT) distribution. We draw an auxiliary random variable as
\begin{equation} \label{eq:crt}
(\ell_{gj}|-) \sim \mbox{CRT}(y_{gj},r_j).
\end{equation}
Then update the cell-dependent dispersion by employing the gamma-Poisson conjugacy
\begin{equation} \label{eq:rji}
(r_j | -) \sim \mbox{Gamma} \Big( a_0+ \sum_g \ell_{gj}, \frac{1}{\gamma - \sum_g \ln(1-p_{gj})} \Big).
\end{equation}

\noindent {\bf Sampling $\boldsymbol{\beta}_g^{(1)}$ and $\boldsymbol{\beta}_j^{(2)}$.\quad} To infer the regression coefficients, we adopt the Polya-Gamma (PG) data augmentation technique~\citep{LGNB_ICML2012,polson2013bayesian}. Denote $\omega_{gj}$ as a random variable drawn from the PG distribution as $\omega_{gj} \sim \text{PG}(y_{gj}+r_j,0)$. 
The likelihood of $\psi_{gj}$ defined in Equation~(1) of the main text can be expressed as 
\begin{align}
\mathcal{L}(\psi_{gj}) &\propto \frac{(e^{\psi_{gj}})^{y_{gj}}}{(1+e^{\psi_{gj}})^{y_{gj}+r_j}} \nonumber\\
& \propto \exp\Big(\frac{y_{gj}-r_j}{2} \psi_{gj}\Big) \cosh^{(y_{gj}+r_j)}(\psi_{gj}^2/2) \nonumber \\
& \propto \exp\Big(\frac{y_{gj}-r_j}{2} \psi_{gj}\Big) \mathbb{E}_{\omega_{gj}}[\exp(- \omega_{gj} \psi_{gj}^2/2)].
\label{eq:lik}
\end{align}
Exploiting the exponential tilting of the PG distribution in \cite{polson2013bayesian}, we draw 
$\omega_{gj}$ as 
\begin{equation}\label{eq:omega}
(\omega_{gj}|-) \sim \text{PG}(y_{gj}+r_j,\psi_{gj}).
\end{equation}
Given the values of the auxiliary variables $\omega_{gj}$ for $j=1,...,J$ and the prior in Equation~(4) of the main text, the conditional posterior of $\boldsymbol{\beta}_g^{(1)}$ can be updated as 
\begin{equation}
(\boldsymbol{\beta}_g^{(1)}|-) \sim \mbox{Normal}(\mu_g^{(\beta^{(1)})},\Sigma_g^{(\beta^{(1)})}),
\label{eq:beta}
\end{equation}
in which $\Sigma_g^{(\beta^{(1)})} = \Big( \mbox{diag}(\alpha_1,...,\alpha_N)+\sum_{j} \omega_{gj} \boldsymbol{x}^{(1)}_j (\boldsymbol{x}_j^{(1)})^T \Big)^{-1}$ and $\mu_g^{(\beta^{(1)})} =$ $\Sigma_g^{(\beta^{(1)})} \Big[ \sum_{j} \big(\frac{y_{gj}-r_j}{2} - \omega_{gj} (\sum_{m} x_{mg}^{(2)} \beta_{jm}^{(2)} + \boldsymbol{\phi}_g^T\boldsymbol{\theta}_j) \big) \boldsymbol{x}^{(1)}_j \Big]$. \\

A similar procedure can be followed to derive the conditional updates for cell-level regression coefficients as
\begin{equation}
(\boldsymbol{\beta}_j^{(2)}|-) \sim \mbox{Normal}(\mu_j^{(\beta^{(2)})},\Sigma_j^{(\beta^{(2)})}),
\label{eq:delta}
\end{equation}
in which $\Sigma_j^{(\beta^{(2)})} = \Big( \mbox{diag}(\eta_1,...,\eta_M)+\sum_{g} \omega_{gj} \boldsymbol{x}^{(2)}_g (\boldsymbol{x}_g^{(2)})^T \Big)^{-1}$ and $\mu_j^{(\beta^{(2)})} =\quad$ $\Sigma_j^{(\beta^{(2)})} \Big[ \sum_{g} \big(\frac{y_{gj}-r_j}{2} - \omega_{gj} (\sum_{n} x_{nj}^{(1)} \beta_{gn}^{(1)} + \boldsymbol{\phi}_g^T\boldsymbol{\theta}_j) \big) \boldsymbol{x}^{(2)}_g \Big]$. 

\noindent {\bf Sampling $\boldsymbol{\phi}_g$ and $\boldsymbol{\theta}_j$.\quad} Using the likelihood function in (\ref{eq:lik}) and the priors defined in Equation (2) of the main text, we can derive closed-form updating steps for factor loading and score parameters. More specifically, the full conditional for factor loading $\boldsymbol{\phi}_v$ is a normal distribution:
\begin{equation}
(\boldsymbol{\phi}_g|-) \sim \mbox{Normal}(\mu_g^{(\phi)},\Sigma_g^{(\phi)}),
\label{eq:phi}
\end{equation}
where $\Sigma_g^{(\phi)} = \Big( I_K + \sum_{j} \omega_{gj} \boldsymbol{\theta}_j \boldsymbol{\theta}_j^T \Big)^{-1}$ and $\mu_g^{(\phi)} = \Sigma_g^{(\phi)} \Big[ \sum_{j} \big(\frac{y_{gj}-r_j}{2} - \omega_{gj} (\sum_{n} x_{nj}^{(1)} \beta_{gn}^{(1)} + \sum_{m} x_{mg}^{(2)} \beta_{jm}^{(2)}) \big) \boldsymbol{\theta}_j \Big]$. 

The full conditional for factor score $\boldsymbol{\theta}_j$ is also a normal distribution:
\begin{equation}
(\boldsymbol{\theta}_j|-) \sim \mbox{Normal}(\mu_j^{(\theta)},\Sigma_j^{(\theta)}),
\label{eq:theta}
\end{equation}
where $\Sigma_j^{(\theta)} = \Big( \mbox{diag}(\gamma_1,...,\gamma_K) + \sum_{g} \omega_{gj} \boldsymbol{\phi}_g \boldsymbol{\phi}_g^T \Big)^{-1}$ and $\mu_j^{(\theta)} = \Sigma_j^{(\theta)} \Big[ \sum_{g} \big(\frac{y_{gj}-r_j}{2} - \omega_{gj} (\sum_{n} x_{nj}^{(1)} \beta_{gn}^{(1)} + \sum_{m} x_{mg}^{(2)} \beta_{jm}^{(2)})  \big) \boldsymbol{\phi}_g \Big]$.

\noindent {\bf Sampling $\boldsymbol{\alpha}_n$ and $\boldsymbol{\eta}_m$.\quad} The precision parameters of the normal distributions in Equation~(4) of the main text can be updated using the normal-gamma conjugacy:
\begin{eqnarray}
	\alpha_n &\sim& \mbox{Gamma} \big( \alpha_0+G/2, \frac{1}{\eta_0 + \sum_{g=1}^{G} (\beta_{gn}^{(1)})^2/2} \big). \nonumber\\
	\eta_m &\sim& \mbox{Gamma} \big( \alpha_0+J/2, \frac{1}{\eta_0 + \sum_{j=1}^{J} (\beta_{jm}^{(2)})^2/2} \big). 
	\label{eq:hyp}
\end{eqnarray}


\noindent {\bf Sampling $\boldsymbol{\gamma}_k$.\quad} Similar as $\alpha_n$ and $\eta_m$, the precision parameter of the normal distributions in Equation~(2) of the main text can be updated as:
\begin{eqnarray}
	\gamma_k &\sim& \mbox{Gamma} \big( e_0+J/2, \frac{1}{f_0 + \sum_{j=1}^{J} (\theta_{jk})^2/2} \big).
	\label{eq:hyp2}
\end{eqnarray}

\noindent {\bf Sampling $\boldsymbol{\nu}$.\quad} Finally, the rate of the gamma distribution for $r_j$ can be updated using the gamma-gamma conjugacy with respect to the rate parameter:
\begin{equation}
	\nu \sim \mbox{Gamma} \big( e_0(1+J), \frac{1}{f_0 + \sum_{j=1}^{J} r_j} \big).
\end{equation}


\subsection*{A.2 ZINB-WaVE count data simulation}

ZINB-WaVE models the gene count $n_{ij}$ as a random variable following a ZINB distribution with parameters $\mu_{ij}$, $\theta_{ij}$ and $\pi_{ij}$
\begin{equation}
	n_{ij} \sim \mbox{ZINB} \big(n_{ij};\mu_{ij},\theta_{ij},\pi_{ij}) = \pi_{ij}\delta_0(n_{ij}) + (1 -\pi_{ij}) \mbox{NB} (n_{ij};\mu_{ij},\theta_{ij}).
\end{equation}
Parameters $\mu_{ij}$, $\theta_{ij}$ and $\pi_{ij}$ assume the following regression model:
\begin{equation}
	\ln(\mu_{ij})  
	= (X\beta_{\mu} + (V\gamma_{\mu})^T + W\alpha_{\mu} + O_{\mu})_{ij};
	\label{eq:zinb_mu}
\end{equation}
\begin{equation}
	\mbox{logit}(\pi_{ij})  
	= (X\beta_{\pi} + (V\gamma_{\pi})^T + W\alpha_{\pi} + O_{\pi})_{ij};
	\label{eq:zinb_pi}
\end{equation}
\begin{equation}
    \ln(\theta_{ij}) = \zeta_{j}.
\end{equation}
In order to change the percentage of inflated (technical) zeros and real (biological) zeros coming from the NB distribution we change the parameters $\gamma_{\pi}$ and $\gamma_{\mu}$ to obtain the desired percentages of zeros coming from either equation (\ref{eq:zinb_mu}) or (\ref{eq:zinb_pi}). 

To simulate the cell clustering structures, a mixture of K(=3)-variate normal distributions with three components is fitted to the inferred $W$ from the real-world data. For each simulated datasets, the low-rank matrix $W$ is generated from the corresponding K-variate normal distribution.

\subsection*{A.3 Additional experimental results with simulated data}

We provide the average evaluation metric values with their standard deviations for all the experiments with the simulated data. Tables~\ref{Tab:EHEHE} and~\ref{Tab:delogfold} provide the detailed differential expression results based on the area under ROC curves~(AUC-ROC) with the hGNB simulated data, corresponding to Figure~1 in the main text. Table~\ref{Tab:EHEHE2} provides the detailed clustering results based on the average silhouette width (ASW) for the hGNB simulated data as visualized in the left panel of Figure~2. In Table~\ref{Tab:clusZINB}, we have the average ASW values with the standard deviations for the clustering results with the data simulated by the ZINB-WaVE model as shown in the right panel of Figure 2 in the main text. 

\subsection*{A.4 Additional results with real-world scRNA-seq data}

\begin{figure*}[]
    \centering
    
        \includegraphics[width=0.7\textwidth,keepaspectratio]{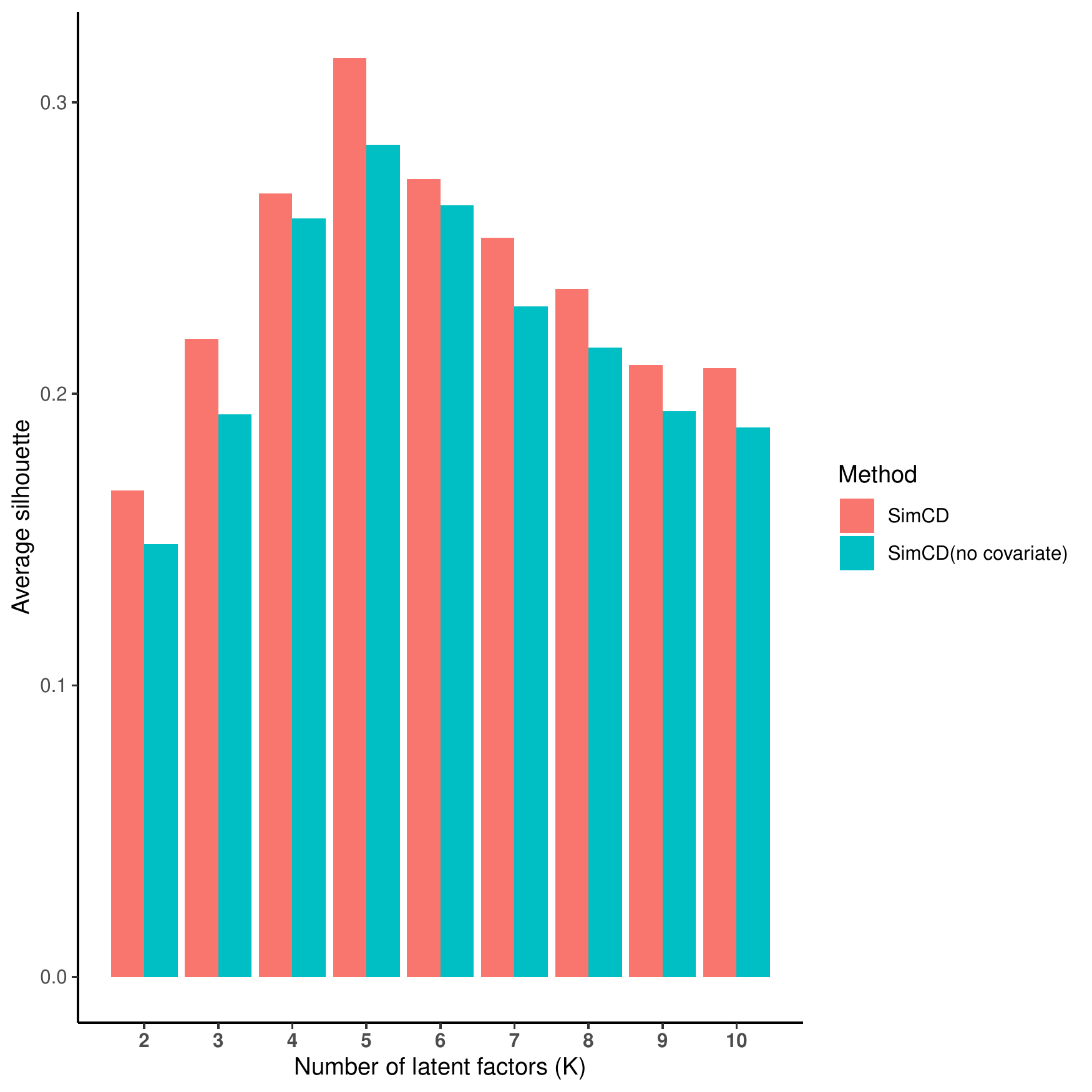}
        \caption{Comparison of clustering ASW on the CORTEX dataset with different numbers of latent factors (K) for the ablation study of including cell-level covariates in SimCD.}
        \label{Fig:K_effect}

\end{figure*}

We provide the average evaluation metric values with their standard deviations (in case of multiple runs) for all the experiments with the studied real-world scRNA-seq data. Table~\ref{Tab:asw_real} provides the detailed clustering results with the studied real-world data, corresponding to Figure 3e in the main text. Moreover, Table~\ref{Tab:depbmc} presents the detailed differential expression analysis results with the PBMC data, specifically between B cells and dendritic cells and between CD$4^+$ and CD$8^+$ cells. This table corresponds to the Figures 4a and b in the main text. In Figure~\ref{Fig:K_effect}, we show the ablation study for the clustering performance of SimCD with and without including cell-level covariates (SimCD vs. SimCD (no covariate)) across different values of the latent space dimension ($K$ in the figure), which is the only hyper-parameter in SimCD that needs to be tuned. As the figure illustrates, SimCD has the best clustering performance when $K=5$ on the Cortex dataset. This figure also highlights the fact that incorporating additional biological covariates like age and sex to the SimCD model can improve its performance, suggesting the need for scRNA-seq data analysis tools that employ cell- and gene-level covariates in their models if possible. 

Tables~\ref{Tab:charac-pbmc} and ~\ref{Tab:charac-mouse} provide the characteristics of cells present in the PBMC and mouse hypothalamus datasets respectively. 


\subsection*{A.5 Gene set enrichment analysis of mouse hypothalamus data}

Table~\ref{Tab:go} shows the top ten enriched GO terms associated with the DE genes detected by SimCD in mouse hypothalamic neuronal subtypes~(Table~\ref{Tab:charac-mouse}). To have a fine-resolution GO enrichment analysis, we evaluate the results based on high-level GO terms in all three categories (biological process (BP), molecular function (MF), and cellular component (CC)). In \cite{shih2012identifying} the authors defined information content (IC) of a GO term $\mbox{g}$ by $\mbox{IC(g)} = -\mbox{log}(|\mbox{g}|/|\mbox{root}|)$, where ``root'' is the corresponding GO category of the GO term $\mbox{g}$. Any GO term with IC $>2$ is considered as a high-level GO term \citep{shih2012identifying}. Table~\ref{Tab:goic} gives the list of enriched high-level GO terms with $\mbox{P-value} < 5 \times 10^{-5}$. It is clear that the top high-level GO terms agree with the current biological understanding of response to food deprivation.

We also provide the top ten enriched GO terms associated with the DE genes detected by DESingle and DESeq2 in Tables~\ref{Tab:go_DEsignle} and \ref{Tab:go_deseq2} respectively.




\begin{table*}[!h]
\centering
\caption{{AUC-ROC of DE analyses with the hGNB simulated data for different zero fractions }
\label{Tab:EHEHE}}
\resizebox{\columnwidth}{!}{
{\begin{tabular}{@{}c|ccccc@{}}
\toprule {\bf ZF} & {\bf SimCD} & {\bf DESeq2} & {\bf DESingle} & {\bf scVI} & {\bf sigEMD}\\\hline \hline
{\bf 20\%} &  {\bf 0.9660 $\pm$ 0.0027} & 0.9332 $\pm$ 0.0104 & 0.9112 $\pm$ 0.0053 & 0.9028 $\pm$ 0.0108 & 0.8501 $\pm$ 0.0453 \\
{\bf 40\%} & {\bf 0.9468 $\pm$ 0.0030} & 0.8997 $\pm$ 0.0078 & 0.8791 $\pm$ 0.0059 & 0.8697 $\pm$ 0.0048 & 0.8316 $\pm$ 0.0116 \\ 
{\bf 60\%} & {\bf 0.9182 $\pm$ 0.0065} & 0.8271 $\pm$ 0.0360 & 0.7991 $\pm$ 0.0300 & 0.7981 $\pm$ 0.0119 & 0.7558 $\pm$ 0.0207 \\
{\bf 80\%} & {\bf 0.8319 $\pm$ 0.0183 } & 0.7263 $\pm$ 0.0143 & 0.7037 $\pm$ 0.0135 & 0.6648 $\pm$ 0.0113 & 0.6384 $\pm$ 0.0134 \\
\hline
\end{tabular}}{}
}
\end{table*}

\begin{table*}[!h]
\centering
\caption{{AUC-ROC of DE analyses with the hGNB simulated data for different true log2-fold changes}
\label{Tab:delogfold}}\vspace{2mm}
\resizebox{\columnwidth}{!}{
{\begin{tabular}{@{}c|ccccc@{}}
\toprule {\bf Abs(Log2-Fold Change)} & {\bf SimCD} & {\bf DESeq2} & {\bf DESingle} & {\bf scVI} & {\bf sigEMD}\\\hline \hline
{\bf 0.485} &  {\bf 0.9196 $\pm$ 0.0079} & 0.8196 $\pm$ 0.0270 & 0.7697 $\pm$ 0.0230 & 0.7663 $\pm$ 0.0334 & 0.7224 $\pm$ 0.0256 \\
{\bf 0.678} & {\bf 0.9529 $\pm$ 0.0049} & 0.8888 $\pm$ 0.0131 & 0.8522 $\pm$ 0.0109 & 0.8503 $\pm$ 0.0131 & 0.8041 $\pm$ 0.0197 \\
{\bf 0.848} & {\bf 0.9660 $\pm$ 0.0027} & 0.9332 $\pm$ 0.0104 & 0.9112 $\pm$ 0.0053 & 0.9028 $\pm$ 0.0108 & 0.8501 $\pm$ 0.0453 \\
{\bf 1.000} & {\bf 0.9666 $\pm$ 0.0038 } & 0.9469 $\pm$ 0.0067 & 0.9313 $\pm$ 0.0077 & 0.9162 $\pm$ 0.0097 & 0.8835 $\pm$ 0.0093 \\
\hline
\end{tabular}}{}
}
\end{table*}

\begin{table*}[!h]
\centering
\caption{{Clustering ASW with the hGNB simulated data for different zero fractions}
\label{Tab:EHEHE2}}
\resizebox{.75\columnwidth}{!}{
{\begin{tabular}{@{}c|ccc@{}}
\toprule {\bf ZF} & {\bf SimCD} & {\bf ZINB-WaVE} & {\bf scVI} \\\hline \hline

{\bf 20\%} &  {\bf 0.5313 $\pm$ 0.0127} & 0.4821 $\pm$ 0.0104 & 0.3152 $\pm$ 0.0183  \\
{\bf 40\%} & {\bf 0.5152 $\pm$ 0.0132} & 0.4467 $\pm$ 0.0178 & 0.2734 $\pm$ 0.0259 \\
{\bf 60\%} & {\bf 0.4756 $\pm$ 0.0165} & 0.3984 $\pm$ 0.0243 & 0.2214 $\pm$ 0.0213  \\
{\bf 80\%} & {\bf 0.4354 $\pm$ 0.0183 } & 0.3419 $\pm$ 0.0143 & 0.1597 $\pm$ 0.0195 \\\hline
\end{tabular}}{}
}
\end{table*}

\begin{table*}[!t]
\centering
\caption{{Clustering ASW with the ZINB-WaVE simulated data for different zero fraction (ZF) levels.}
\label{Tab:clusZINB}}
\resizebox{.85\columnwidth}{!}{
{\begin{tabular}{@{}c|c|ccc@{}}
\toprule {\bf ZF} & {\bf Ratio}  & {\bf SimCD} & {\bf ZINB-WaVE} & {\bf scVI} \\\hline \hline
{} & {1} & {\bf 0.4867 $\pm$ 0.0031} &  0.4456 $\pm$ 0.0036 & 0.4612 $\pm$ 0.0044 \\
{\bf 20\%} & {5} & {\bf 0.3988 $\pm$ 0.0018}  & 0.3851 $\pm$ 0.0023  & 0.3768 $\pm$ 0.0034 \\
{} & {10} & {\bf 0.2422 $\pm$ 0.0016}  &  0.2403 $\pm$ 0.0012 & 0.2113 $\pm$ 0.0021\\ \hline
{} & {1} & {\bf 0.4233 $\pm$ 0.0025}   &  0.3894 $\pm$ 0.0066 & 0.3934 $\pm$ 0.0027\\
{\bf 40\%} & {5} & {\bf 0.3504 $\pm$ 0.0042}   &  0.3386 $\pm$ 0.0074 & 0.3362 $\pm$ 0.0028\\
{} & {10} & {\bf 0.1926  $\pm$ 0.0024}  &  0.1856  $\pm$ 0.0028 &  0.1669 $\pm$ 0.0051 \\ \hline
{} & {1} & {\bf 0.3329 $\pm$ 0.0042}  &  0.3064 $\pm$ 0.0082 & 0.2684 $\pm$ 0.0076  \\
{\bf 60\%} & {5} & {\bf 0.2384 $\pm$ 0.0063 }&  0.2159 $\pm$ 0.0091 & 0.1996 $\pm$ 0.0058 \\
{} & {10} & 0.1189 $\pm$ 0.0018  & {\bf 0.1208 $\pm$ 0.0029} & 0.0925 $\pm$ 0.0074 \\ \hline
{} & {1} & {\bf 0.1812 $\pm$ 0.0028}    &  0.1592 $\pm$ 0.0025 & 0.1226 $\pm$ 0.0062 \\
{\bf 80\%} & {5} & {\bf 0.1113 $\pm$ 0.0025}  & 0.1096 $\pm$ 0.0029 & 0.0674 $\pm$ 0.0057 \\
{} & {10} & 0.0492 $\pm$ 0.0056 & {\bf 0.0581 $\pm$ 0.0055} &  0.0276 $\pm$ 0.0077 \\ \hline
\end{tabular}}{}
}
\end{table*}

\begin{table*}[!t]
\centering
\caption{{Clustering ASW of selected methods on three real-world scRNA-seq count data}
\label{Tab:asw_real}}
\resizebox{.9\columnwidth}{!}{
{\begin{tabular}{@{}c|cccc@{}}
\toprule {\bf Dataset} & {\bf SimCD} & {\bf ZINB-WaVE} & {\bf scVI} & {\bf SimCD (no covariate)}\\\hline \hline

{\bf CORTEX} &  {\bf 0.3153} & 0.2436  & 0.2977 & 0.2854  \\
{\bf PBMC4k} & {\bf 0.4699} & 0.4420 & 0.4163 & - \\
{\bf Hypothalamus} & {\bf 0.2347} & 0.1309  & 0.0678  &  -
\\\hline
\end{tabular}}{}
}
\end{table*}

\begin{table*}[!t]
\centering
\caption{{AUC-ROC of DE analyses across selected cell clusters by selected methods on the PBMC dataset}
\label{Tab:depbmc}}\vspace{2mm}
\resizebox{\columnwidth}{!}{
{\begin{tabular}{@{}c|ccccc@{}}
\toprule {\bf Cell groups} & {\bf SimCD} & {\bf DESeq2} & {\bf DESingle} & {\bf scVI} & {\bf sigEMD}\\\hline \hline
{\bf B cells} &   &  &  &   &  \\
{\bf Vs} &  {\bf 0.7517 $\pm$ 0.0104} & 0.6508 $\pm$ 0.0112 & 0.6405 $\pm$ 0.0128 & 0.7125 $\pm$ 0.0071 & 0.6299 $\pm$ 0.0131 \\ {\bf Dendritic cells} &   &  &  &   &  \\
\hline
{\bf CD$4^+$ cells} &   &  &  &   &  \\
{\bf Vs} & {\bf 0.8720 $\pm$ 0.0155} & 0.5482 $\pm$ 0.0257 & 0.8027 $\pm$ 0.0154 & 0.8140 $\pm$ 0.017 & 0.5316 $\pm$ 0.0156 \\
{\bf CD$8^+$ cells} &   &  &  &   &  \\
\hline
\end{tabular}}{}
}
\end{table*}

\begin{table*}[!t]
\centering
\caption{{Characteristics of cell types present in PBMC dataset.\label{Tab:charac-pbmc}}}
\resizebox{.85\columnwidth}{!}{
{\begin{tabular}{@{}cccc@{}}
\toprule {\bf Cluster name} & {\bf \# Cells in PBMC4k} & {\bf \# Cells in PBMC8k} & {\bf \# Total cells}  \\\hline \hline 
B cells &  554  &  1071  &  1625  \\
CD1$4^+$ Monocytes &  742  &  1495  &  2237\\
CD4 T cells &  1647  &  3377  &  5024\\
CD8 T cells &  499  &  953  &  1452\\
Dendritic cells  &  128  &  211  &  339\\
FCGR3$A^+$ Monocytes  &  125  &  226  &  351\\
Megakaryocytes &  25  &  63  &  88\\
NK cells &  172  &  287  &  459\\
Other &  117  &  347  &  464\\
\hline
Total  &  4009  &  8030  &  12039\\
\end{tabular}}
}%
\end{table*}

\begin{table*}[!t]
\centering
\caption{{Characteristics of mouse hypothalamic neuronal subtypes that are responsive to food deprivation reported in \cite{mousehypo2017}.\label{Tab:charac-mouse}}}
\resizebox{.8\columnwidth}{!}{
{\begin{tabular}{@{}cccc@{}}
\toprule {\bf Cluster name} & {\bf \# Normal cells} & {\bf \# Food-deprived cells} & {\bf \# Total cells}  \\\hline \hline 
GABA 1 &  1  &  9  &  10  \\
GABA 11 &  8  &  17  &  25\\
GABA 15 &  31  &  37  &  68\\
GABA 18 &  4  &  19  &  23\\
Glu 5 &  36  &  47  &  83\\
Glu 8 &  20  &  17  &  37\\
Glu 12 &  4  &  13  &  17\\
\hline
Total  &  104  &  159  &  263\\
\end{tabular}}
}%
\end{table*}

\begin{table*}[!t]
\centering
\caption{{Top ten enriched GO terms associated with DE genes detected by SimCD in mouse hypothalamic neuronal subtypes}
\label{Tab:go}}
\resizebox{.75\columnwidth}{!}{
{\begin{tabular}{@{}cccc@{}}
\toprule {\bf GO ID} & {\bf Ontology} & {\bf Description} & {\bf P-value} \\\hline \hline

GO:0031982 &  CC  &  Vesicle  &  5.0e-07 \\
GO:0031988 &  CC  &  Membrane-bounded vesicle  & 1.5e-06 \\
GO:0097458 &  CC  & Neuron part &  5.9e-06  \\
GO:0005184 &  MF  &  Neuropeptide hormone activity  & 9.9e-06 \\
GO:0043523 &  BP  &  Regulation of neuron apoptotic process  &  1.1e-05 \\
GO:0032879 &  BP  &  Regulation of localization  &  1.2e-05 \\
GO:0060341 &  BP  &  Regulation of cellular localization  &  1.3e-05 \\
GO:0065008 &  BP  &  Regulation of biological quality  &  2.4e-05 \\
GO:0051402 &  BP  &  Neuron apoptotic process  &  	2.6e-05 \\
GO:0043025 &  CC  &  neuronal cell body  &  	2.7e-05 \\
\hline
\end{tabular}}{}
}
\end{table*}

\begin{table*}[!t]
\centering
\caption{{Top ten enriched GO terms associated with DE genes detected by DESingle in mouse hypothalamic neuronal subtypes}
\label{Tab:go_DEsignle}}
\resizebox{.75\columnwidth}{!}{
{\begin{tabular}{@{}cccc@{}}
\toprule {\bf GO ID} & {\bf Ontology} & {\bf Description} & {\bf P-value} \\\hline \hline

GO:0031982 &  CC  &  Vesicle  &  4.0e-12 \\
GO:0031988 &  CC  &  Membrane-bounded vesicle  & 1.3e-11 \\
GO:0098793 &  CC  & presynapse &  4.2e-11  \\
GO:0008021 &  CC  &  synaptic vesicle  & 1.3e-10 \\
GO:0070062 &  CC  &  extracellular exosome  &  3.2e-08 \\
GO:0065010 &  CC  &  extracellular membrane-bounded organelle  &  3.5e-08 \\
GO:1903561 &  CC  &  extracellular vesicle  &  3.7e-08 \\
GO:0043230 &  CC  &  extracellular organelle  &  3.9e-08 \\
GO:0031410 &  CC  &  cytoplasmic vesicle  &  	1.5e-07 \\
GO:0044433 &  CC  &  cytoplasmic vesicle part  &  	2.8e-07 \\
\hline
\end{tabular}}
}
\end{table*}

\begin{table*}[!t]
\centering
\caption{{Top ten enriched GO terms associated with DE genes detected by DESeq2 in mouse hypothalamic neuronal subtypes}
\label{Tab:go_deseq2}}
\resizebox{.75\columnwidth}{!}{
{\begin{tabular}{@{}cccc@{}}
\toprule {\bf GO ID} & {\bf Ontology} & {\bf Description} & {\bf P-value} \\\hline \hline

GO:0031982 &  CC  &  Vesicle  &  7.6e-07 \\
GO:0070062 &  CC  &  extracellular exosome  & 1.4e-06 \\
GO:0065010 &  CC  & extracellular membrane-bounded organelle &  1.5e-06  \\
GO:1903561 &  CC  &  extracellular vesicle  & 1.6e-06 \\
GO:0043230 &  CC  &  extracellular organelle  &  1.6e-06 \\
GO:0031988 &  CC  &  membrane-bounded vesicle  &  5.3e-06 \\
GO:0044421 &  CC  &  extracellular region part  &  5.0e-05 \\
GO:0046185 &  BP  &  aldehyde catabolic process  &  9.6e-05 \\
GO:0015992 &  BP  &  proton transport  &  	1.1e-04 \\
GO:0006818 &  BP  &  hydrogen transport  &  	1.2e-04 \\
\hline
\end{tabular}}{}
}
\end{table*}

\begin{table*}[!t]
\centering
\caption{{Enriched high-level GO terms associated with DE genes detected by SimCD in mouse hypothalamic neuronal subtypes}
\label{Tab:goic}}
\resizebox{.75\columnwidth}{!}{
{\begin{tabular}{@{}ccccc@{}}
\toprule {\bf GO ID} & {\bf Ontology} & {\bf Description} & {\bf P-value}  & {\bf IC} \\\hline \hline
GO:0097458 &  CC  & Neuron part &  5.9e-06  & 2.5\\
GO:0005184 &  MF  &  Neuropeptide hormone activity  & 9.9e-06 & 6.5 \\
GO:0043523 &  BP  &  Regulation of neuron apoptotic process  &  1.1e-05 & 4.2\\
GO:0060341 &  BP  &  Regulation of cellular localization  &  1.3e-05 & 2.6\\
GO:0051402 &  BP  &  Neuron apoptotic process  &  	2.6e-05 & 4.2\\
GO:0043025 &  CC  &  neuronal cell body  &  	2.7e-05 & 3.3\\
GO:0051050 &  BP  &  positive regulation of transport  &  	3.3e-05 & 2.9\\
GO:0023061 &  BP  &  signal release  &  	3.8e-05 & 3.6\\
GO:0051049 &  BP  &  regulation of transport  &  	3.9e-05 & 2.3\\
\hline
\end{tabular}}{}
}
\end{table*}

\clearpage

%
\bibliography{references}

\begin{thebibliography}{}

\bibitem[Appleyard(2003)Appleyard]{Gal2003}
Appleyard, S. (2003).
\newblock Appetite regulation, neuronal control.
\newblock {\em Encyclopedia of Hormones\/}, pages 171--179.

\bibitem[Boluki {\em et~al.}(2020)Boluki, Qian, and Dougherty]{Shahin}
Boluki, S.  {\em et~al.} (2020).
\newblock Optimal {B}ayesian supervised domain adaptation for {RNA} sequencing
  data.
\newblock {\em under review\/}.

\bibitem[Brogan {\em et~al.}(1997)Brogan, Fife, Conley, Giustina, and
  Wehrenberg]{Ghrh1997}
Brogan, R.  {\em et~al.} (1997).
\newblock Effects of food deprivation on the gh axis: immunocytochemical and
  molecular analysis.
\newblock {\em Neuroendocrinology\/}, {\bf 65}(2), 129--35.

\bibitem[Campbell and Yau(2018)Campbell and Yau]{campbell2018uncovering}
Campbell, K.~R. and Yau, C. (2018).
\newblock Uncovering pseudotemporal trajectories with covariates from single
  cell and bulk expression data.
\newblock {\em Nature communications\/}, {\bf 9}(1), 2442.

\bibitem[Chen {\em et~al.}(2017)Chen, Wu, Jiang, and Zhang]{mousehypo2017}
Chen, R.  {\em et~al.} (2017).
\newblock Single-cell rna-seq reveals hypothalamic cell diversity.
\newblock {\em Cell Rep\/}, {\bf 18}(13), 3227--3241.

\bibitem[Choi {\em et~al.}(2020)Choi, Chen, Skelly, and
  Churchill]{choiK2020zerosorigin}
Choi, K.  {\em et~al.} (2020).
\newblock Bayesian model selection reveals biological origins of zero inflation
  in single-cell transcriptomics.
\newblock {\em Genome Biol\/}, {\bf 21}(1), doi: 10.1186/s13059--020--02103--2.

\bibitem[Cole {\em et~al.}(2017)Cole, Risso, and Wagner]{scone2017}
Cole, M.~B.  {\em et~al.} (2017).
\newblock Performance assessment and selection of normalization procedures for
  single-cell rna-seq.
\newblock {\em bioRxiv\/}.

\bibitem[Dadaneh {\em et~al.}(2018)Dadaneh, Qian, and Zhou]{dadaneh2018bnp}
Dadaneh, S.~Z.  {\em et~al.} (2018).
\newblock {BNP-seq: B}ayesian nonparametric differential expression analysis of
  sequencing count data.
\newblock {\em Journal of the American Statistical Association\/}, {\bf
  113}(521), 81--94.

\bibitem[Dadaneh {\em et~al.}(2020)Dadaneh, de~Figueiredo, Sze, Zhou, and
  Qian]{dadaneh2020bayesian}
Dadaneh, S.~Z.  {\em et~al.} (2020).
\newblock Bayesian gamma-negative binomial modeling of single-cell rna
  sequencing data.
\newblock {\em BMC genomics\/}, {\bf 21}(9), 1--10.

\bibitem[Gong {\em et~al.}(2018)Gong, Kwak, Pota, Koyano-Nakagawa, and
  Garry]{DrImpute2018}
Gong, W.  {\em et~al.} (2018).
\newblock Drimpute: imputing dropout events in single cell rna sequencing data.
\newblock {\em BMC Bioinformatics\/}, {\bf 19}(220).

\bibitem[Hinton and Roweis(2003)Hinton and Roweis]{Hinton_Roweis_2003}
Hinton, G. and Roweis, S. (2003).
\newblock Stochastic neighbor embedding.
\newblock In S.~T. S~Becker and K.~Obermayer, editors, {\em Advances in neural
  information processing systems\/}, volume~15, pages 833--840.

\bibitem[Janowski {\em et~al.}(1993)Janowski, Ling, Giustina, and
  Wehrenberg]{Ghrh1993}
Janowski, B.  {\em et~al.} (1993).
\newblock Hypothalamic regulation of growth hormone secretion during food
  deprivation in the rat.
\newblock {\em Life Sci\/}, {\bf 52}(11), 981--7.

\bibitem[Jiang {\em et~al.}(2015)Jiang, Modise, Helm, Jensen, and
  Good]{Jiang2015}
Jiang, H.  {\em et~al.} (2015).
\newblock Characterization of the hypothalamic transcriptome in response to
  food deprivation reveals global changes in long noncoding rna, and cell cycle
  response genes.
\newblock {\em Genes Nutr.}, {\bf 10}(6), 48.

\bibitem[Johnson {\em et~al.}(2005)Johnson, Kemp, and
  Kotz]{johnson2005univariate}
Johnson, N.~L.  {\em et~al.} (2005).
\newblock {\em Univariate discrete distributions\/}, volume 444.
\newblock John Wiley \& Sons.

\bibitem[Klami {\em et~al.}(2013)Klami, Virtanen, and Kaski]{klami2013bayesian}
Klami, A.  {\em et~al.} (2013).
\newblock Bayesian canonical correlation analysis.
\newblock {\em Journal of Machine Learning Research\/}, {\bf 14}(Apr),
  965--1003.

\bibitem[Kyrkouli {\em et~al.}(1986)Kyrkouli, Stanley, and Leibowitz]{Gal1986}
Kyrkouli, S.  {\em et~al.} (1986).
\newblock Galanin: stimulation of feeding induced by medial hypothalamic
  injection of this novel peptide.
\newblock {\em Eur J Pharmacol\/}, {\bf 122}(1).

\bibitem[Leek {\em et~al.}(2012)Leek, Johnson, Parker, Jaffe, and
  Storey]{sva2012}
Leek, J.~T.  {\em et~al.} (2012).
\newblock The sva package for removing batch effects and other unwanted
  variation in high-throughput experiments.
\newblock {\em Bioinformatics\/}, {\bf 28}(6).

\bibitem[Lopez {\em et~al.}(2018)Lopez, Regier, Cole, Jordan, and
  Yosef]{lopez2018deep}
Lopez, R.  {\em et~al.} (2018).
\newblock Deep generative modeling for single-cell transcriptomics.
\newblock {\em Nature methods\/}, {\bf 15}(12), 1053--1058.

\bibitem[Love {\em et~al.}(2014)Love, Huber, and Anders]{deseq2}
Love, M.~I.  {\em et~al.} (2014).
\newblock moderated estimation of fold change and dispersion for rna-deq data
  with deseq2.
\newblock {\em Genome Biology\/}, {\bf 15}(550).

\bibitem[Lytal {\em et~al.}(2020)Lytal, Ran, and An]{Lytal2020}
Lytal, N.  {\em et~al.} (2020).
\newblock Normalization methods on single-cell rna-seq data: An empirical
  survey.
\newblock {\em Front Genet\/}, {\bf 11}, 41.

\bibitem[Miao {\em et~al.}(2018)Miao, Deng, Wang, and Zhang]{DEsingle}
Miao, Z.  {\em et~al.} (2018).
\newblock Desingle for detecting three types of differential expression in
  single-cell rna-seq data.
\newblock {\em Bioinformatics\/}, {\bf 34}(18), 3223--3224.

\bibitem[Mickelsen {\em et~al.}(2019)Mickelsen, Bolisetty, Chimileski, and
  et~al]{LHAhypo2019}
Mickelsen, L.  {\em et~al.} (2019).
\newblock Single-cell transcriptomic analysis of the lateral hypothalamic area
  reveals molecularly distinct populations of inhibitory and excitatory
  neurons.
\newblock {\em Nat Neurosci\/}, {\bf 22}, 642--656.

\bibitem[Mou {\em et~al.}(2020)Mou, Deng, Gu, Pawitan, and Vu]{Mou2020}
Mou, T.  {\em et~al.} (2020).
\newblock Reproducibility of methods to detect differentially expressed genes
  from single-cell rna sequencing.
\newblock {\em Frontiers in Genetics\/}, {\bf 10}, 1331.

\bibitem[{Noorbala} {\em et~al.}(2019){Noorbala}, {Niyakan}, and {Mahdi
  Alavi}]{tDMI}
{Noorbala}, L.  {\em et~al.} (2019).
\newblock Development of phase congruency to estimate the direction of maximum
  information (tdmi) in images with straight line segments.
\newblock In {\em 2019 27th Iranian Conference on Electrical Engineering
  (ICEE)\/}, pages 1413--1419.

\bibitem[Perraudeau {\em et~al.}(2017)Perraudeau, Risso, and
  Street]{Perraudeau2017}
Perraudeau, F.  {\em et~al.} (2017).
\newblock Bioconductor workflow for single-cell rna sequencing: Normalization,
  dimensionality reduction, clustering, and lineage inference.
\newblock {\em F1000Research\/}, {\bf 6}(1158).

\bibitem[Polson {\em et~al.}(2013)Polson, Scott, and
  Windle]{polson2013bayesian}
Polson, N.~G.  {\em et~al.} (2013).
\newblock Bayesian inference for logistic models using p{\'o}lya--gamma latent
  variables.
\newblock {\em Journal of the American statistical Association\/}, {\bf
  108}(504), 1339--1349.

\bibitem[Purdom {\em et~al.}(2017)Purdom, Risso, Helm, Jensen, and
  Good]{cluster2017}
Purdom, E.  {\em et~al.} (2017).
\newblock clusterexperiment: Compare clusterings for single-cell sequencing.
\newblock {\em R package version\/}, {\bf 1}(0).

\bibitem[Qualls-Creekmore {\em et~al.}(2017)Qualls-Creekmore, Yu, Francois,
  Hoang, Huesing, Bruce-Keller, Burk, Berthoud, Morrison, and
  Münzberg]{Gal2017}
Qualls-Creekmore, E.  {\em et~al.} (2017).
\newblock Galanin-expressing gaba neurons in the lateral hypothalamus modulate
  food reward and noncompulsive locomotion.
\newblock {\em J Neurosci\/}, {\bf 37}(25), 6053--6065.

\bibitem[Risso {\em et~al.}(2011)Risso, Schwartz, Sherlock, and
  Dudoit]{risso2011gc}
Risso, D.  {\em et~al.} (2011).
\newblock Gc-content normalization for rna-seq data.
\newblock {\em BMC bioinformatics\/}, {\bf 12}(1), 480.

\bibitem[Risso {\em et~al.}(2018)Risso, Perraudeau, Gribkova, Dudoit, and
  Vert]{zinbwave2018}
Risso, D.  {\em et~al.} (2018).
\newblock A general and flexible method for signal extraction from single-cell
  rna-seq data.
\newblock {\em Nature Communications\/}, {\bf 9}(284).

\bibitem[Schroeder and Leinninger(2018)Schroeder and Leinninger]{Nts2018}
Schroeder, L. and Leinninger, G. (2018).
\newblock Hypothalamic regulation of growth hormone secretion during food
  deprivation in the rat.
\newblock {\em Biochim Biophys Acta Mol Basis Dis\/}, {\bf 1864}(3), 900--916.

\bibitem[Shalek {\em et~al.}(2014)Shalek, Satija, Shuga, Trombetta, Gennert,
  and et~al.]{paracrine2014}
Shalek, A.~K.  {\em et~al.} (2014).
\newblock Single-cell rna-seq reveals dynamic paracrine control of cellular
  variation.
\newblock {\em Nature\/}, {\bf 510}, 363--369.

\bibitem[Shih and Parthasarathy(2012)Shih and
  Parthasarathy]{shih2012identifying}
Shih, Y.-K. and Parthasarathy, S. (2012).
\newblock Identifying functional modules in interaction networks through
  overlapping markov clustering.
\newblock {\em Bioinformatics\/}, {\bf 28}(18), i473--i479.

\bibitem[Van~den Berge {\em et~al.}(2018)Van~den Berge, Perraudeau, Soneson,
  Love, Risso, and et~al.]{vandenberge2018}
Van~den Berge, K.  {\em et~al.} (2018).
\newblock Observation weights unlock bulk rna-seq tools for zero inflation and
  single-cell applications.
\newblock {\em Genome Biology\/}, {\bf 19}(24).

\bibitem[Van~den Berge {\em et~al.}(2020)Van~den Berge, De~Bezieux, Street,
  Saelens, Cannoodt, Saeys, Dudoit, and Clement]{van2020trajectory}
Van~den Berge, K.  {\em et~al.} (2020).
\newblock Trajectory-based differential expression analysis for single-cell
  sequencing data.
\newblock {\em Nature communications\/}, {\bf 11}(1), 1--13.

\bibitem[Wang and Nabavi(2018)Wang and Nabavi]{sigEMD}
Wang, T. and Nabavi, S. (2018).
\newblock Sigemd: A powerful method for differential gene expression analysis
  in single-cell rna sequencing data.
\newblock {\em Methods\/}, {\bf 145}, 25--32.

\bibitem[Woodworth {\em et~al.}(2017)Woodworth, Beekly, Batchelor, Bugescu,
  Perez-Bonilla, Schroeder, and Leinninger]{Nts2017}
Woodworth, H.  {\em et~al.} (2017).
\newblock Lateral hypothalamic neurotensin neurons orchestrate dual weight loss
  behaviors via distinct mechanisms.
\newblock {\em Cell Rep\/}, {\bf 21}(11), 3116--3128.

\bibitem[Wu and Ma(2020)Wu and Ma]{jointDRclustering}
Wu, W. and Ma, X. (2020).
\newblock Joint learning dimension reduction and clustering of single-cell
  rna-sequencing data.
\newblock {\em Bioinformatics\/}, {\bf 36}(12), 3825–3832.

\bibitem[Zeisel {\em et~al.}(2015)Zeisel, Munoz-Manchado, Codeluppi, and
  et~al.]{Zeisel2015}
Zeisel, A.  {\em et~al.} (2015).
\newblock Cell types in the mouse cortex and hippocampus revealed by
  single-cell rna-seq.
\newblock {\em Science\/}, {\bf 347}, 1138--1142.

\bibitem[Zheng {\em et~al.}(2017)Zheng, Terry, and Belgrader]{pbmc2017}
Zheng, G.  {\em et~al.} (2017).
\newblock Massively parallel digital transcriptional profiling of single cells.
\newblock {\em Nature communications\/}, {\bf 8}(14049).

\bibitem[Zhou and Carin(2015)Zhou and Carin]{zhou2015negative}
Zhou, M. and Carin, L. (2015).
\newblock Negative binomial process count and mixture modeling.
\newblock {\em IEEE Transactions on Pattern Analysis and Machine
  Intelligence\/}, {\bf 37}(2), 307--320.

\bibitem[Zhou {\em et~al.}(2012)Zhou, Li, Dunson, and Carin]{LGNB_ICML2012}
Zhou, M.  {\em et~al.} (2012).
\newblock Lognormal and gamma mixed negative binomial regression.
\newblock In {\em International Conference on Machine Learning\/}, volume 2012,
  page 1343.

\bibitem[Zyprych-Walczak {\em et~al.}(2015)Zyprych-Walczak, Szabelska,
  Handschuh, and et~al.]{Zyprych2015}
Zyprych-Walczak, J.  {\em et~al.} (2015).
\newblock The impact pf normalization methods on rna-seq data analysis.
\newblock {\em BioMed Research International\/}.

\end{thebibliography}

 \bibliographystyle{natbib}


\end{document}